# Demonstration of highly scaled AlScN ferroelectric diode memory with storage density > 100 Mbit/mm$^2$

Zekun Hu[1], Hyunmin Cho[1], Rajeev Kumar Rai[2], Kefei Bao[1], Yinuo Zhang[1], Zhaosen Qu[3], Yunfei He[1], Yaoyang Ji[1], Chloe Leblanc[1], Kwan-Ho Kim[1], Zirun Han[1], Zhen Qiu[1], Xingyu Du[1], Eric A. Stach[2], Roy Olsson[1], Deep Jariwala[1,*]

[1]Department of Electrical and System Engineering, University of Pennsylvania, Philadelphia, Pennsylvania 19104, United States of America

[2]Department of Materials Science and Engineering, University of Pennsylvania, Philadelphia, Pennsylvania 19104, United States of America

[3]Department of Electrical and Computer Engineering, Johns Hopkins University, Baltimore, Maryland 21218, United States of America

[*]Correspondence should be addressed: dmj@seas.upenn.edu (D. J.)

# Abstract

Wurtzite nitride ferroelectric materials have emerged as promising candidates for next-generation memory applications due to their exceptional polarization properties and compatibility with conventional semiconductor processing techniques. Here, we demonstrate the first successful areal scaling of Aluminum Scandium Nitride (AlScN) ferroelectric diode (FeDiode) memory down to 40 nm device diameters while maintaining ON/OFF > 60. Using a 20 nm thick $Al_{0.64}Sc_{0.36}N$ ferroelectric layer, we evaluate both metal-insulator-ferroelectric-metal (MIFM) and metal-ferroelectric-metal (MFM) architectures for scaled resistive memory devices. Our scaled devices exhibit an enhanced breakdown-to-coercive field ratio exceeding 2.6 due to increased breakdown field. The MIFM devices demonstrate stable 3-bit non-volatile multistate behavior with clearly distinguishable resistance states and retention exceeding $4\times10^4$ seconds at 85 °C. By achieving more than a million-fold areal scaling with enhanced performance metrics, this work establishes AlScN-based FeDiode memory as a highly promising platform for non-volatile storage with potential for direct integration into CMOS technology.

The development of high-density, non-volatile memory (NVM) compatible with existing semiconductor logic technology is essential for advancing computing systems beyond current limitations[1]. While volatile memory such as SRAM and DRAM as well as traditional NVM technologies such as NOR flash memory face significant scaling challenges[2], emerging ferroelectric materials provide a promising path forward due to their intrinsic non-volatility, fast switching capabilities, and monolithic integration potential. Aluminum Scandium Nitride (AlScN) stands out as a particularly promising nitride ferroelectric material for memory applications. AlScN exhibits several critical advantages over other ferroelectric candidates, including a remarkably high remanent polarization exceeding 100 μC/cm², substantially higher than oxide-based ferroelectrics such as $HfO_2$[3, 4]. For practical implementation, AlScN thin films can be deposited at temperatures below 350°C, ensuring compatibility with back-end-of-line (BEOL) CMOS processes and enabling seamless integration with existing semiconductor manufacturing[5].

The ferroelectric diode (FeDiode) device architecture leverages the inherent ferroelectric polarization dependent leakage current properties of AlScN to create memory devices with built-in rectification, eliminating the need for additional selector elements typically required in high-density memory arrays[5, 6]. However, a critical question remains unaddressed in the literature: Can wurtzite FeDiodes maintain consistent properties such as current density, ON/OFF ratio, and retention when scaled to dimensions necessary for competitive areal memory densities? While theoretical calculations suggest favorable scaling potential[7], experimental demonstration of functional, scaled AlScN FeDiodes has been lacking.

In this work, we present the first demonstration of area-scalable $Al_{0.64}Sc_{0.36}N$ (hereafter referred as AlScN)-based FeDiode memory with devices successfully fabricated and characterized from 50 um down to 40 nm diameters (more than a million fold scaling in device area). We systematically investigate two device architectures: a metal-insulator-ferroelectric-metal (MIFM) structure incorporating a 4 nm $HfO_2$ insulating layer and a metal-ferroelectric-metal (MFM) structure without an interfacial insulator. We observed an increase in breakdown field and consistent current density as the device dimensions were reduced. For devices scaled to 50 nm diameter, we achieve an ultra-high breakdown field to coercive field ratio exceeding 2.6, significantly improving reliability and operating range.

Beyond binary switching, we demonstrate stable three-bit multistate operation in our scaled FeDiodes. This multistate capability is maintained with clear distinction between resistance states and retention exceeding $5\times10^5$ seconds with projected stability beyond 10 years. By achieving 40 nm scaling while enhancing critical performance metrics, this work establishes AlScN-based FeDiode memory as a highly promising candidate for next-generation non-volatile storage with direct integration potential for advanced computing architectures.

A schematic depiction of the Au (120 nm)/ Ti (30 nm)/ $HfO_2$ (4 nm)/ AlScN (20 nm)/ Al (111) (50 nm) MIFM FeDiode with a via structure appears in **Figure 1a, d**. The 20 nm thick AlScN film deposits via co-sputtering from separate Al and Sc targets onto an in-situ grown (111)-textured Al layer over a 4-inch sapphire wafer to reduce lattice mismatch. The top surface initially receives another in-situ Al layer to prevent oxidation of the ferroelectric. After wet etching, electron-beam lithography (EBL) defines markers, which are patterned by ebeam evaporation of Ti/Au and subsequent lift-off, followed by the deposition of a 100 nm $SiO_2$ layer, which is then dry etched to form the via. A 4-nm $HfO_2$ insulating layer is then deposited via atomic layer deposition (ALD) for the MIFM structure, after which the Ti/Au top electrode evaporates and patterns using lift-off. A detailed fabrication process appears in SI Figure S1. The fabrication of MFM FeDiodes follows the same procedure, except without the ALD insulator deposition step.

The via structures are characterized using scanning electron microscopy (SEM), as shown in **Figure 1c**. SEM images confirm successful scaling down to 50 nm. **Figure 1e** shows a representative cross-sectional scanning transmission electron microscopy (STEM) image of a 4 μm diameter MIFM device. STEM analysis confirms the uniformity of the insulating and ferroelectric layers, with the measured thicknesses of AlScN and $HfO_2$ being 20 nm and 4 nm, respectively. The high-resolution STEM image shown in **Figure 1f** reveals a sharp and uniform interface without any dead layer between AlScN and Al. A (0002) oriented AlScN has been grown over (111) Al. A 2–3 nm oxidized region was also observed between AlScN and $HfO_2$. This oxidized layer, while potentially beneficial in suppressing leakage through passivation, may also contribute to device-to-device variability and influence long-term endurance and reliability. **Figure 1g** shows a Fast Fourier Transform (FFT) from the AlScN/Al interface, confirming the crystallographic orientation relationship between AlScN and Al. The AlScN thin film adopts a wurtzite crystal structure, resulting in a measured c/a ratio of 1.56. **Figure 1h** provides an energy-dispersive X-ray spectroscopy (EDS) analysis of the diode via region, verifying the elemental distribution of respective elements in corresponding thin films.

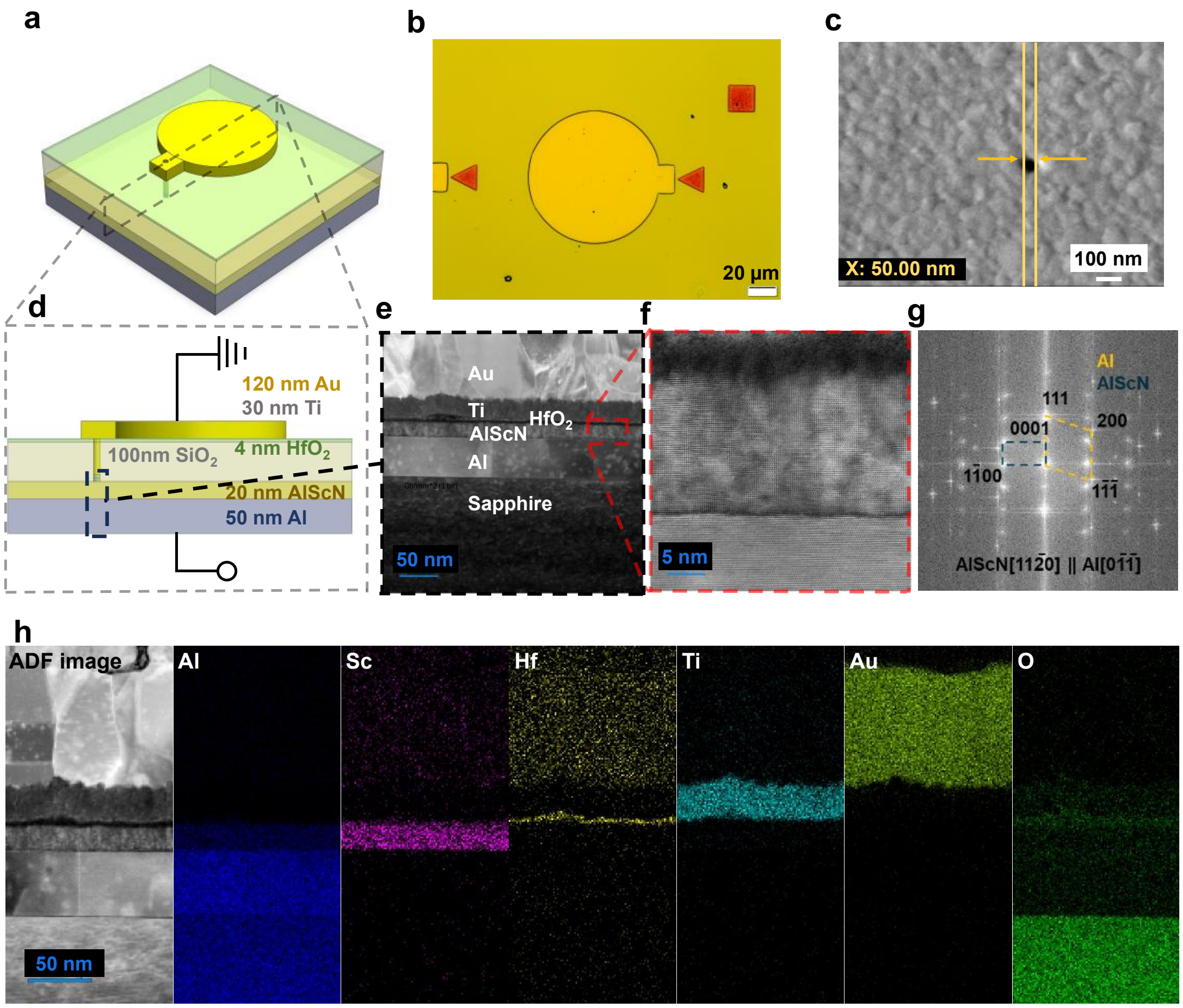

**Figure 1.** Structural and morphological characterization of the fabricated MIFM device.(a) Schematic illustration of the device layout. (b) Optical image of a fabricated device with a 4 µm diameter. (c) SEM image showing the surface morphology and the structural details of a 50 nm diameter device. (d) Cross-sectional schematic of the device architecture, highlighting the layer composition. (e) Cross-sectional STEM image of the via, showing the layer stack including the AlScN ferroelectric layer and $HfO_2$ insulating layer. (f) High-resolution TEM image of the AlScN/Al interface, illustrating the material quality and interface characteristics. (g) FFT pattern of image shown in (e) confirming the crystallographic orientation of the AlScN layer. (h) EDS mapping of the TEM cross-section, displaying the elemental distribution of aluminum, scandium, hafnium, titanium, gold, and oxygen across the device structure.

To systematically evaluate the impact of FeDiode scaling on memory characteristics, we investigate its electrical performance, including switching voltage, leakage behavior, and breakdown reliability. **Figure 2** presents the direct circuit (DC) electrical measurements of both MIFM and MFM structures across a wide range of device diameters, demonstrating their non-volatile memory operation and scalability. Quasi-DC (0.01 Hz) current-voltage (I–V) hysteresis measurements are performed with diameters ranging from 50 μm to 50 nm, as shown in **Figure 2a–c**. The applied maximum voltages of 12 V for MIFM and 11 V for MFM ensure complete ferroelectric switching.

The I–V characteristics of MIFM devices (**Figure 2a**) show clear current responses across all voltage ranges for large devices, with a stable ON-OFF ratio maintained for voltages above 5 V. **Figure 2b** presents the current density as a function of voltage, showing strong consistency across different device sizes at voltages above 5 V. The peak current densities align well, confirming robust fabrication accuracy.

For MFM devices, the current density characteristics (**Figure 2c**) similarly exhibit consistent scaling trends, with all device sizes demonstrating matching current densities. However, since MFM devices lack an insulating $HfO_2$ layer, the maximum ON-OFF ratio is reduced, aligning with previous observations[5]. The current characteristics of MFM devices, along with additional DC I–V curves, are provided in SI Figure S4.

As shown in **Figure 2a–c**, the DC I–V characteristics exhibit a clear distinction between the high-resistance state (HRS) and low-resistance state (LRS) before reaching the switching voltage of 7–9 V. As a ferroelectric memory, the FeDiode demonstrates a relatively large memory window compared to HZO-based ferroelectric memory, attributed to AlScN's high coercive field and large remanent polarization[8]. . As shown in Figure 2f, the scaled 40 nm diameter device exhibits robust and uniform performance. Considering its device area of 0.00126 μm² and operation as a 1 bit memory cell, our FeDiode achieves a potential storage density of 225 Mbit/mm², surpassing commercial NOR Flash by more than 20 times.

**Figure 2d** compares the extracted current density between the ON-state and OFF-state for both MIFM and MFM structures. The MFM devices exhibit high uniformity across all sizes, suggesting that the observed variation in MIFM devices originates from the presence of the $HfO_2$ insulating layer. Despite this variability, the insulating layer effectively suppresses OFF-state leakage current by reducing tunneling probability. We verified this by modelling the tunneling current with the Poole-Frenkel model in both the ON and OFF states, as shown in SI (Figure S8 and Note 1). Additionally, the insulating layer enhances the on-state current density, leading to an increased On-Off ratio.

Observing the ON-state across all sizes, which is less susceptible to measurement noise, reveals a slightly higher current density in smaller devices. This behavior is attributed to local perimeter electric field reinforcement[9], which becomes more pronounced as the device scales down. The variation in MFM devices remains negligible, indicating that the ferroelectric properties of AlScN are preserved during scaling down to a 50 nm diameter.

To evaluate high speed write capability, triangular pulses of varying frequencies up to 1 MHz were applied to program the device, followed by a 5 V DC read to eliminate frequency induced noise. As shown in **Figure 2e**, the read current increases progressively with pulse amplitude and saturates once the device is fully programmed.

To directly confirm ferroelectric switching behavior, we perform pulse switching and Positive-Up-Negative-Down (PUND) polarization measurements. Results clearly indicate ferroelectric polarization reversal, with a saturated 2Pr of approximately 220 μC/cm², shown in SI Figure S17.

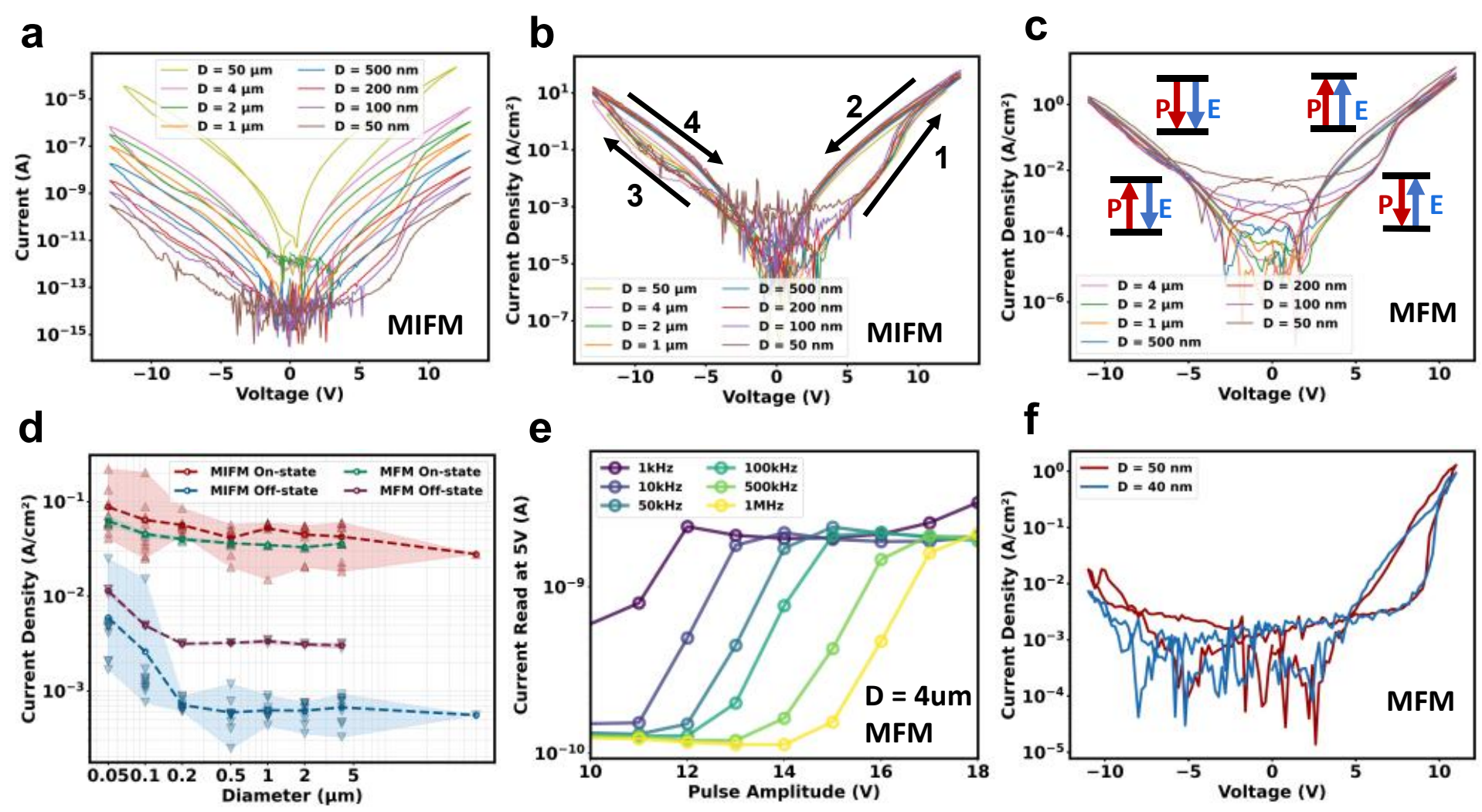

**Figure 2**. DC electrical characterization of MIFM and MFM structures across different device sizes. (a) I-V characteristics of MIFM structures for various diameters. (b) Current density-voltage characteristics of MIFM and MFM structures. (c) I-V characteristics of MFM structures across different diameters. (d) Current density comparison of ON-state and OFF-state for both MIFM and MFM structures, with statistical variations. (e) AC pulse program with DC resistance readout, demonstrating successful switching up to 1 MHz for MFM devices with D = 4 µm. Current was measured at 5 V after each pulse sequence. (f) DC I–V characteristics of a 40 nm device compared to a 50 nm device from the same chip, showing consistent current density, confirming reliable scaling down to sub-50 nm dimensions.

By applying a read voltage of 5 V, the on-state and OFF-state currents were extracted from 20 DC I–V measurements per size and used to determine the ON-OFF ratio, as shown in **Figure 3a**. When comparing the performance of MIFM and MFM structures, the addition of the insulating layer enhances state differentiation, particularly in smaller devices. This observation is consistent with previous reports showing that the insertion of an insulator modulates the electrostatic potential barrier and helps suppress direct tunneling[5].

**Figure 3b** shows a consistent switching voltage for the MFM structure, corresponding to a DC coercive field of approximately 3.6 MV/cm**.** The gradient plot used to determine the switching voltage is shown in SI Figure S5. The varying via hole diameter for the MIFM device varies the growth rate of the $HfO_2$ layer, causing a slightly thinner (and more variable) $HfO_2$ layer for smaller diameter devices and a reduced switching voltage.

The ratio of the dielectric breakdown field to the coercive field is shown in **Figure 3c**. As the device dimensions are reduced, the breakdown field increases while the coercive field decreases, leading to a higher breakdown-to-coercive-field ratio. For MFM devices, the measured ratio aligns with reported values at a 4 μm diameter but exceeds 2.6 when scaled down to 50 nm, demonstrating an improvement over existing reports[4, 10, 11]. This enhancement suggests that device scaling plays a critical role in improving performance metrics, potentially due to increased electric field effects and improved charge modulation at reduced dimensions.

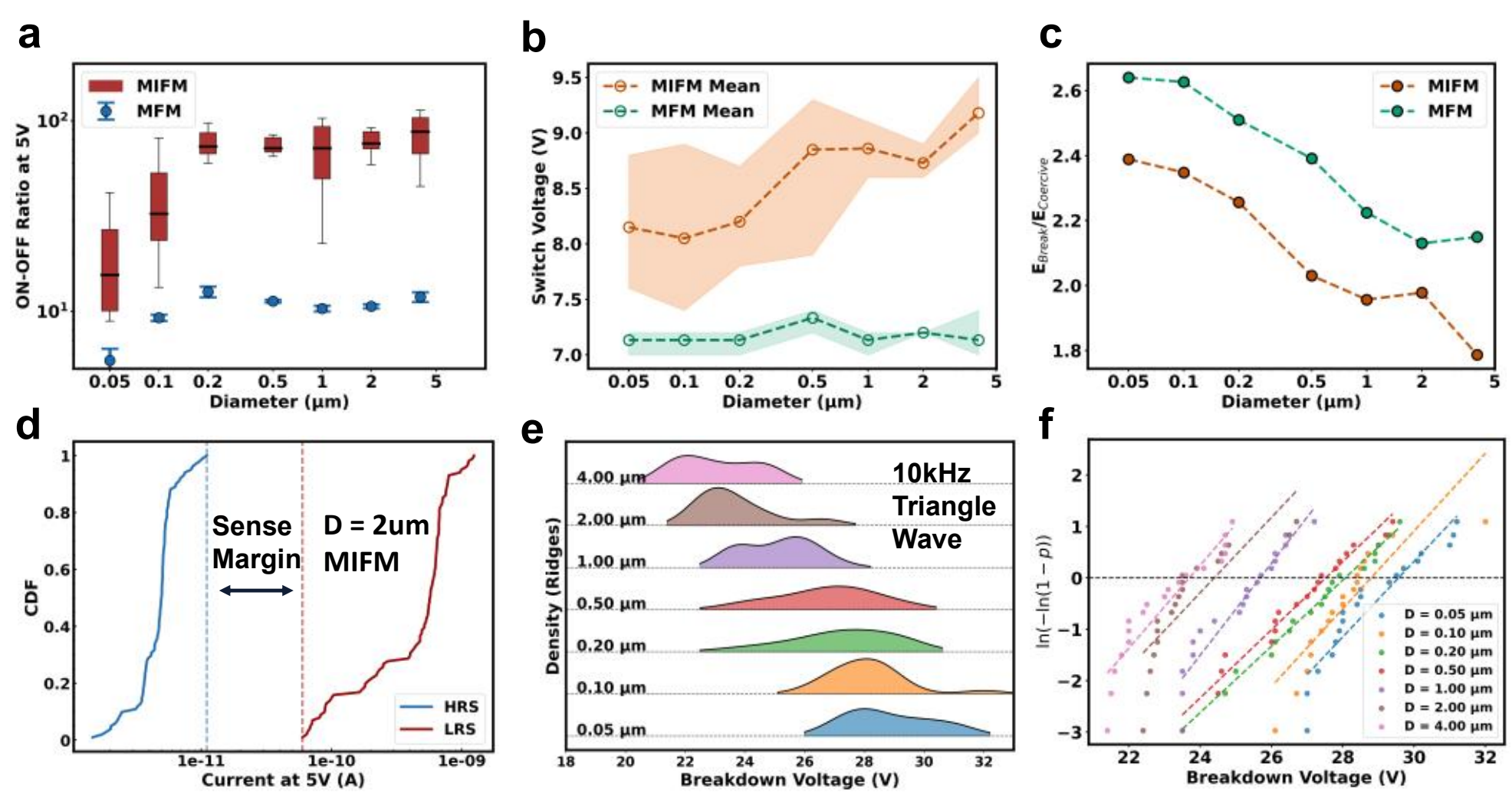

**Figure 3.** Performance comparison of MIFM and MFM structures in terms of ferroelectric field characteristics and breakdown behavior. (a) ON-OFF ratio at 5V across various device sizes. (b) DC ferroelectric switching voltage, illustrating size-dependent trends. (c) Ratio of breakdown field to coercive field. (d) Cumulative distribution function (CDF) of 100 MIFM devices with D = 2 μm, showing distinct separation between HRS and LRS current at 5 V, confirming consistent switching behavior and a reliable sensing margin. (e) MIFM Breakdown voltage distribution for 20 devices of each size under 10 kHz operation. (f) Weibull plot and corresponding linearization to visualize the statistical reliability of MIFM breakdown voltage across different sizes.

The measured 100 MIFM devices with a diameter of 2 μm and exhibit a clear sense margin and excellent uniformity, as shown in **Figure 3d**. Additional efforts toward crossbar two-state verification of selector free operation are presented in SI Figure S21, where the yield and performance will be further optimized in future works.

The breakdown voltages of 20 MIFM devices for each size are presented in **Figure 3e**. A 10 kHz triangular bipolar waveform was applied, with increments of 0.1 V until breakdown occurred. Once a device reaches breakdown, short-circuit behavior and visible metal damage can be observed, as shown in SI Figure S7. A key benefit of scaling down is the increase in breakdown field, consistent with previous studies[11, 12].

The Weibull distribution of breakdown field, shown in **Figure 3f**, provides statistical insight into the reliability of the devices. The results reveal a clear trend where smaller device sizes consistently exhibit higher breakdown voltages. For the 50 nm diameter devices, the electric breakdown field at 63% probability reaches 12.2 MV/cm.

While two-state (1-bit) switching provides a reliable foundation for non-volatile memory applications for simple, long term data storage, the ability to store multiple intermediate states further enhances the storage density and computational potential of FeDiodes. The introduction of an insulating layer in FeDiode enhances its suitability for multistate operation by increasing the ON-OFF ratio and reducing the switching slope of the I-V curves. Unlike conventional memory architectures, FeDiodes offer an efficient approach to processing-in-memory, reducing the need for additional selectors or memristors for complex computations[6, 13].

The multistate behavior of the MIFM device was verified by increasing the program voltage from 7 V to 12 V in 0.1 V steps, as shown in SI Figure S9. As illustrated in **Figure 4a–b**, 3-bit (8 states) multistate behavior is demonstrated for devices larger than 200 nm in diameter, where both the ratio to the OFF-state and the current density increases progressively from the lowest state to the fully programmed state (state 8). The measurement was conducted using DC program voltages, followed by three consecutive DC read operations at 5 V to ensure accuracy and repeatability. Detailed test procedures are provided in SI Figure S10. For sub-200 nm devices, unstable low read signals (SI Figure. S11) were improved using a 7 V read voltage, enabling multistate behavior to 50 nm (SI Figure. S12).

Here, when operating as a multistate device, each programmed state retains its value for over 1000 seconds at room temperature, as shown in **Figure 4c**. The measured 1 μm diameter device exhibit clear and distinguishable differences read continuously at 5V, further confirming that the multistate behavior arises from partial switching of the ferroelectric AlScN layer.

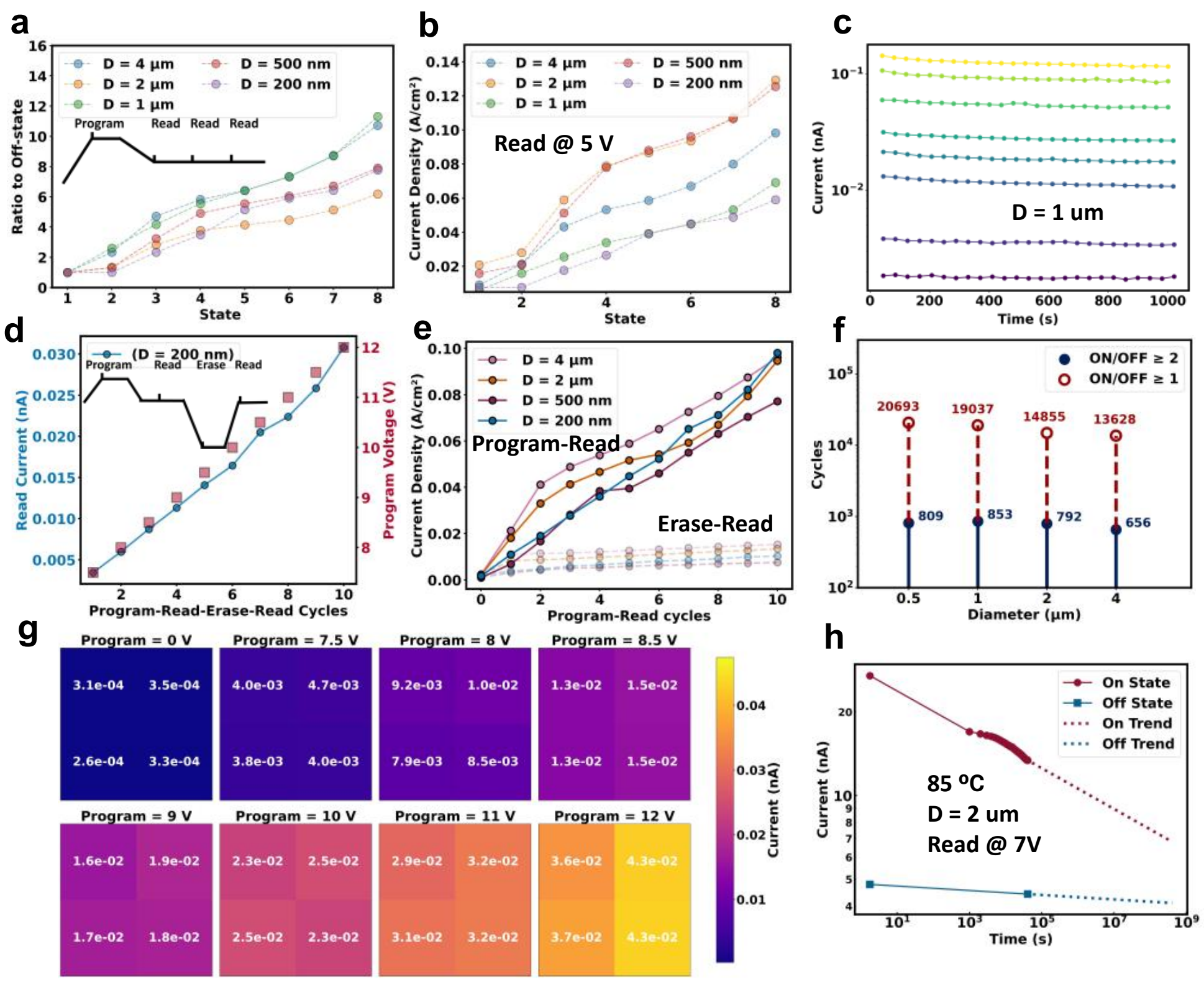

**Figure 4.** Multistate behavior of MIFM devices across various sizes down to D = 200 nm. (a) Ratio to the OFF-state for each programmed state over multiple program-read cycles, with program voltages ranging from 7.5 V to 12 V. (b) Current density of each programmed state, showing distinct multilevel characteristics. (c) Retention behavior of each state over time for a D = 1 μm device, demonstrating data stability. (d) Read current at 5 V for a D = 200 nm device during program-read-erase-read cycles, with erase voltages of -11 V. (e) Current read after program and erase operations for different device sizes, showing consistency across scaling. (f) AC-programmed, DC-read endurance test. Devices maintaining ON/OFF > 2 are marked in blue and ON/OFF > 1 in red. Results show improved endurance with reduced device diameter. (g) Color map representation of the output current at 5 V for four D = 200 nm devices over program-read-erase cycles, visualizing multistate conductance levels as a function of programming voltage. (h) Two-state retention test at 85 °C, illustrating the stability of the ON-state and OFF-state currents over time.

In addition to the increasing programming voltage with different read current responses, a more rigorous approach was applied to demonstrate the reconfigurable multistate memory behavior of the MIFM FeDiode. As shown in **Figure 4d**, a program-read-erase-read sequence was implemented on 200 nm devices to assess its performance across multiple cycles. The detailed test setting is shown in SI Figure S14. The read current gradually increases in a linear manner, indicating a linear conductance variation across different states. This linearity is crucial for low-power neural network computing, enabling efficient analog computation in artificial intelligence hardware[14].

As shown in **Figure 4e**, devices >200 nm show clear, size-independent multistate behavior, confirming ferroelectric AlScN scaling. For smaller devices, 5 V readout is noise-limited; 7 V results are in SI Figure S15. To evaluate device-to-device variation, we test four separate 200 nm devices, with the results shown in Figure 4g. While variation is clearly present across devices, the programmed states remain reasonably distinguishable. To account for this variability conservatively, the storage density projection is based on reliable two-bit operation, ensuring practical scalability and robustness.

Since it is difficult to accurately measure 2Pr in the small devices, the endurance shown in Figure 4f is estimated using an AC programming and DC readout approach. The corresponding data, detailed in SI Figure S18, confirm stable device operation over repeated switching cycles. Notably, we observe a scaling benefit in the form of increased endurance at smaller dimensions, which we attribute to the enhanced breakdown-to-coercive field ratio enabled by dimensional scaling.

We evaluated the stability of two resistance states at elevated temperature (85°C) for over $4\times10^4$ seconds, ensuring long-term data retention. Additionally, the FeDiodes show retention performance comparable to FeFETs[15] with a simpler fabrication process and improved CMOS compatibility.

Finally, we compare and benchmark our FeDiode technology to various other volatile and non-volatile memory technologies in the table below. We demonstrate the scalability of our FeDiode device, achieving a storage density of 225 Mb/mm² at the 40 nm node under a $4F^2$-limited crossbar array architecture, with the potential to reach 625 Mb/mm² for the 20 nm node, while maintaining stable single-bit operation. The storage density is calculated in SI Note 2. Although selector-free operation is already well demonstrated for FeDiode architectures[16], we additionally include a selector-free crossbar feasibility analysis based on our experimentally measured rectification behavior in SI Note 3 to further support the integration potential.

Although the switching voltage is higher than in some emerging memory technologies, it can be further reduced through optimization of ferroelectric thickness, doping concentration, and contact engineering. These improvements can enhance performance while maintaining the high integration density and selector-free operation of the AlScN FeDiode. Its key novelty lies in device scaling with invariant properties—unlike RRAM, PCRAM, or MRAM, which degrade with miniaturization[17].

While the fabrication process remains relatively complex and the ON/OFF ratio can be improved, these limitations are addressable through further material and structural refinement. This study demonstrates meaningful progress toward scalable integration, with ongoing efforts focused on lowering switching voltage, improving multistate stability at nanoscale dimensions, and evaluating endurance and variability at the array level.

**Table1.** Benchmarking between high-density memory technology

| **Memory technology** | **SRAM** | **DRAM** | **Ferroelectric (HZO) DRAM** | **NOR flash** | **MRAM** | **PCRAM** | **RRAM** | **FeDiode** |
|---|---|---|---|---|---|---|---|---|
| **Cell Sstructure** | 6T | 1T-1C | 1T-1C | 1Floating Gate | 1T-1R | 1T-1R | 1T-1R | 1FeDiode (built-in selector) |
| **Cell size** | $>100F^2$ | $>6F^2$ | $>4F^2$ | $>10F^2$ | $>6F^2$ | $>4F^2$ | $>4F^2$ | **$4F^2$** |

| | | | | | | | | | |
|---|---|---|---|---|---|---|---|---|---|
| **Process node (nm)** | F ≈ 20 | F ≈ 30 | F ≈ 24 | F ≈ 100 | F ≈ 100 | F ≈ 100 | F ≈ 90 | F = 40 | F = 20 |
| **Storage density (Gb/$mm^2$)** | 0.038 | 0.3 | 0.225 (per layer) | 0.01 | 0.0125 | 0.05 | 0.02 (per layer) | 0.225 (1 bit) | **0.625 (1 bit)** |
| **Non-Volatility** | No | No | Yes | Yes | Yes | Yes | Yes | Yes | |
| **On/Off Ratio** | N/A | N/A | N/A | 1E4 - 1E5 | ~2.5 | 1E2 | 10 | 1E2 | |
| **Write voltage** | <1 V | <1 V | 1.5 V | >10 V | <0.5 V | <3 V | <3 V | <10 V | |
| **Retention** | N/A | milliseconds to seconds | >10 years | >10 years | >10 years | >10 years | >10 years | >10 years | |
| **References** | [18] | [19] | [20] | [21] | [22] | [23] | [24] | **This work** | |

To conclude, we have demonstrated the first successful scaling of AlScN-based ferroelectric diode memory to 40 nm dimensions, a storage density of 225 Mbit/$mm^2$ for a device area of 0.00126 $\mu m^2$ at 1-bit operation establishing a new paradigm for high-density, non-volatile storage. Our systematic investigation reveals a remarkable size-dependent enhancement in device performance where switching voltage decreases and breakdown field increases as dimensions shrink, resulting in an ultra-high breakdown-to-coercive field ratio of 2.6 at 50 nm. The MIFM architecture, incorporating a thin $HfO_2$ insulating layer, enables stable 3-bit storage with distinct resistance states, demonstrating the potential for high-density multibit storage and compute-in-memory functionality in selector-free crossbar arrays. Our FeDiodes maintain exceptional retention characteristics, with demonstrated stability over $4\times10^4$ seconds at 85 °C and projected retention beyond 10 years. This work highlights the potential of BEOL-compatible AlScN for scalable ferroelectric memory, offering multistate operation and integration readiness for stand-alone, embedded, and compute-in-memory applications.

**Supplementary Information includes:** materials and methods; ferroelectric characterization including PUND and frequency-dependent hysteresis; AC program–DC read retention and endurance measurements; temperature-dependent tests; device-to-device variability and CDF analysis; statistical switching voltage extraction; real crossbar demonstration; and selector-free calculation based on measured data.

# Acknowledgements

We thank Issac Kelvin from the University of Pennsylvania for assistance with data collection. We also acknowledge the support and helpful suggestions from David Barth and the staff at the Quattrone Nanofabrication Facility, University of Pennsylvania.

**Funding:**

D.J. and Z.H. acknowledge primary support from the Office of Naval Research (ONR) Nanoscale Computing and Devices program (N00014-24-1-2131). D.J. also acknowledges partial support from the Air Force Office of Scientific Research (AFOSR) GHz-THz program grant number FA9550-23-1-0391. R.H.O. acknowledges partial support from the Army/ARL via the Collaborative for Hierarchical Agile and Responsive Materials (CHARM) under cooperative agreement W911NF-19-2-0119. A portion of the sample fabrication, assembly and characterization were carried out at the Singh Center for Nanotechnology at the University of Pennsylvania, which is supported by the National Science Foundation (NSF) National Nanotechnology Coordinated Infrastructure Program grant NNCI-1542153. Additional support for the electron microscopy characterization was supported by the NSF through the University of Pennsylvania Materials Research Science and Engineering Center (MRSEC) (DMR-2309043).

**Data availability**

All data are available in the paper and Supplementary Information. Additional data are available from the corresponding authors upon reasonable request.

**Competing interests:**

The authors declare no competing interests.

Supporting information for

# Demonstration of highly scaled AlScN ferroelectric diode memory with storage density > 100 Mbit/mm$^2$

Zekun Hu[1], Hyunmin Cho[1], Rajeev Kumar Rai[2], Kefei Bao[1], Yinuo Zhang[1], Zhaosen Qu[3], Yunfei He[1], Yaoyang Ji[1], Chloe Leblanc[1], Kwan-Ho Kim[1], Zirun Han[1], Zhen Qiu[1], Xingyu Du[1], Eric A. Stach[2], Roy Olsson[1], Deep Jariwala[1,*]

[1]Department of Electrical and System Engineering, University of Pennsylvania, Philadelphia, Pennsylvania 19104, United States of America

[2]Department of Materials Science and Engineering, University of Pennsylvania, Philadelphia, Pennsylvania 19104, United States of America

[3]Department of Electrical and Computer Engineering, Johns Hopkins University, Baltimore, Maryland 21218, United States of America

[*]Correspondence should be addressed: dmj@seas.upenn.edu (D. J.)

**Figure S1-S22**

**Note 1-3**

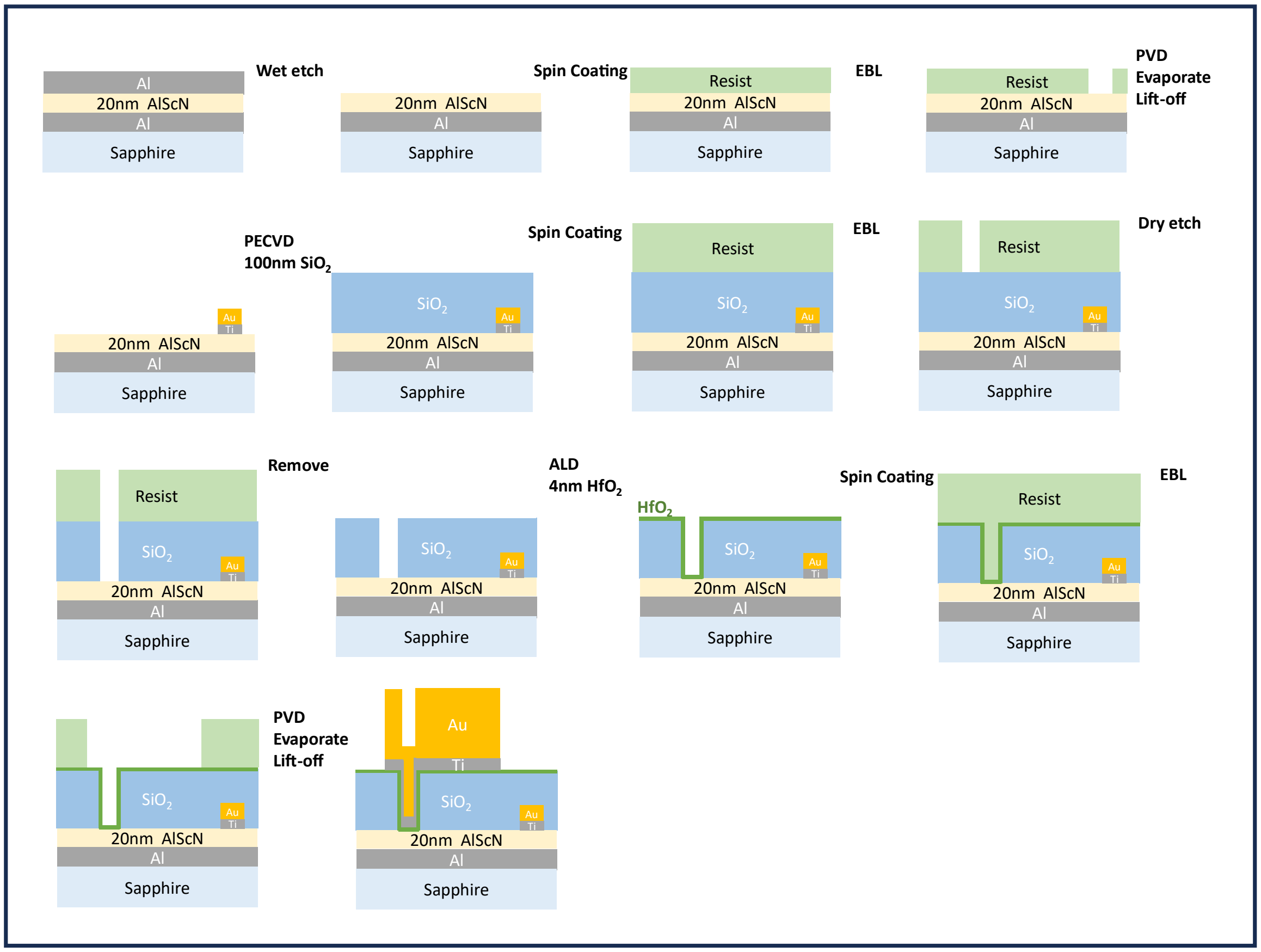

**Figure S1.** Fabrication process flow for MIFM and MFM devices. The process begins with wet etching and spin coating, followed by electron beam lithography for pattern definition. Subsequent steps include dry etching, resist removal, and deposition of dielectric and electrode materials. Plasma-enhanced chemical vapor deposition (PECVD) is used for $SiO_2$ deposition, and atomic layer deposition (ALD) is employed for $HfO_2$ deposition. Thermal evaporation is used for metal deposition, followed by a lift-off process to define the final electrode structure. The schematic illustrates each step in sequence, highlighting material layers and processing techniques.

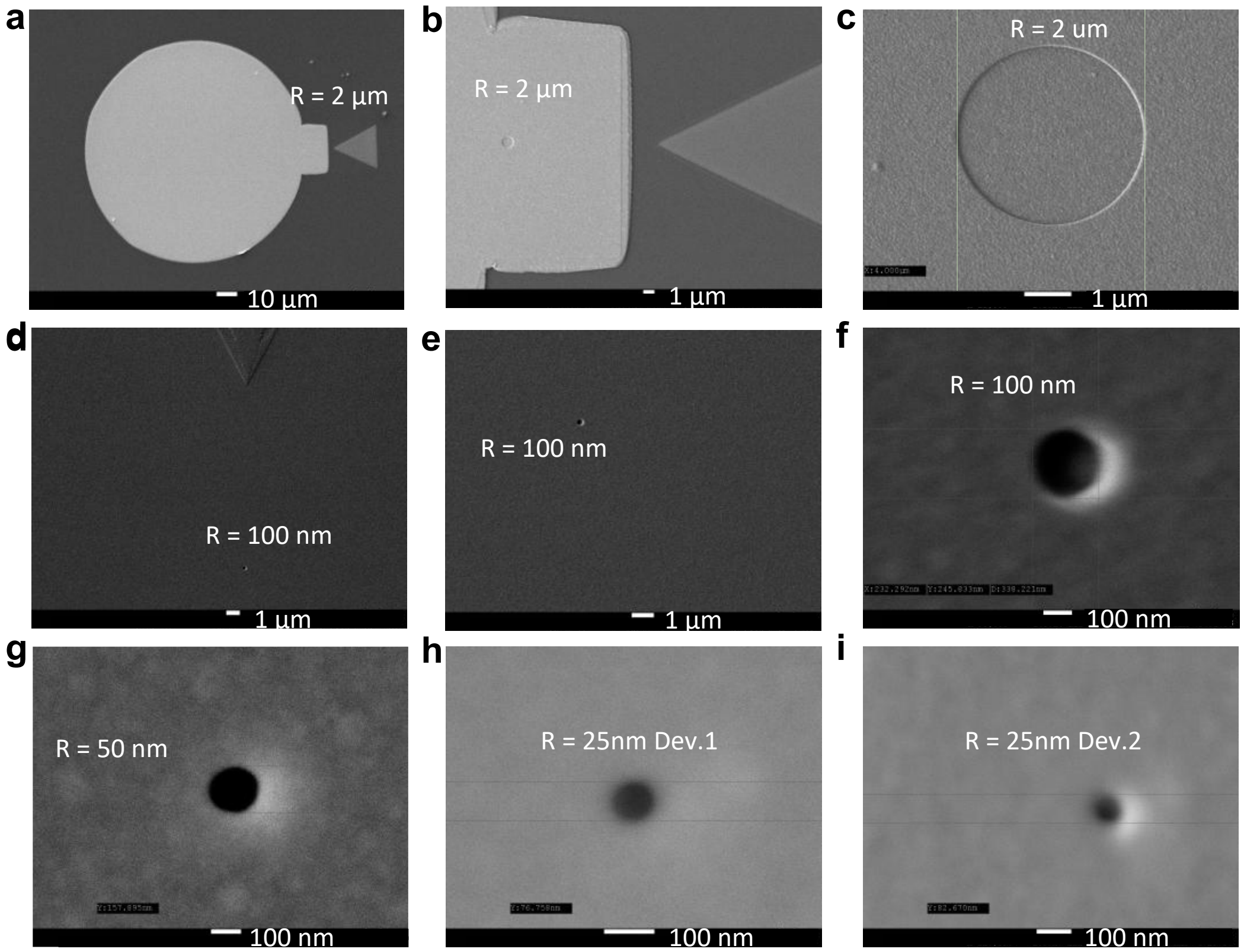

**Figure S2.** SEM images. a-c. After metal deposition. d-i. After etching the $SiO_2$ via.

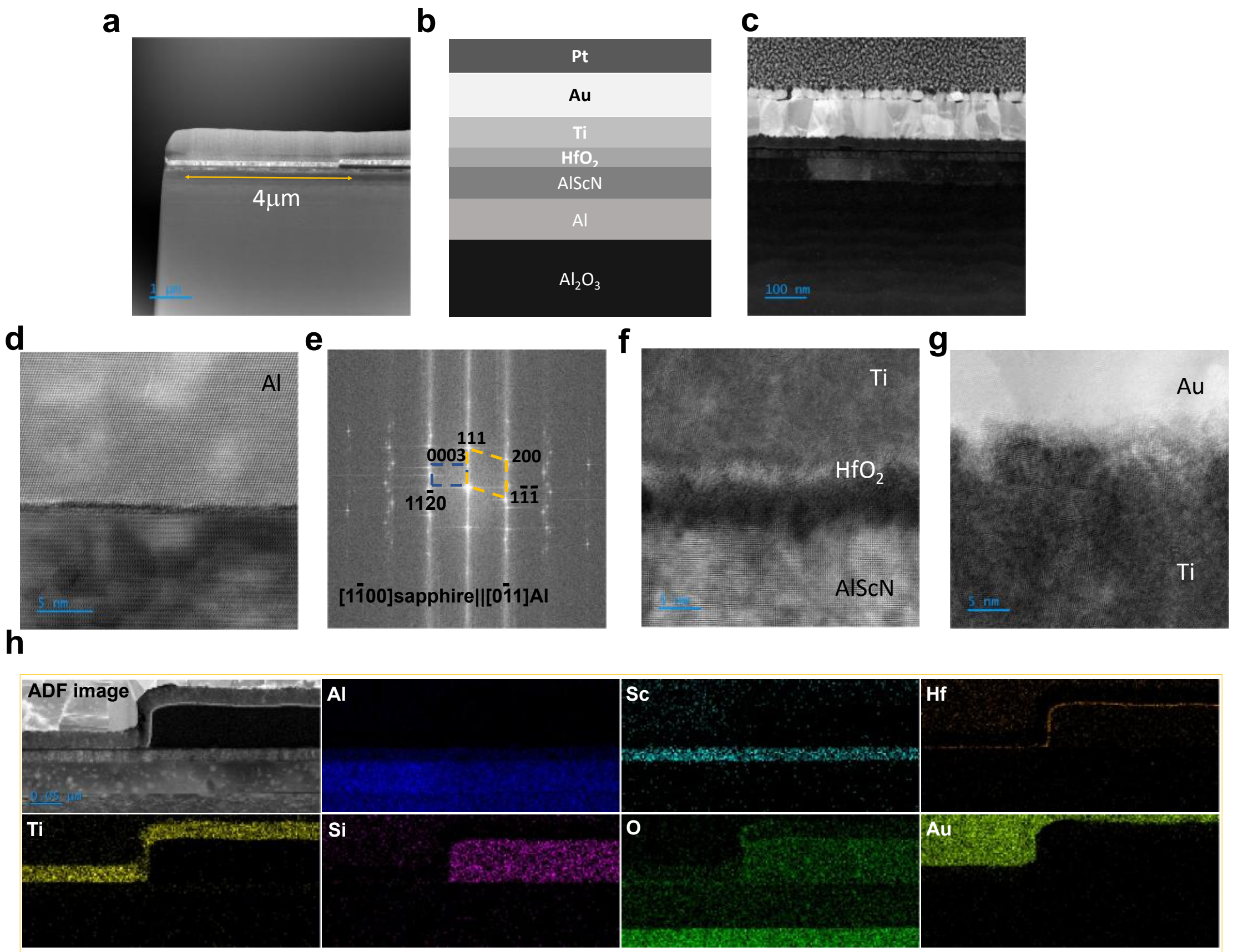

**Figure S3.** STEM Characterization of the FeDiode Structure. (a) Cross-sectional STEM images of a device with a 4 µm diameter, showing distinct material layers. (b) Schematic illustration of the layer stack in the TEM cross-section. (c)Cross-sectional STEM images. (e) EDS of Al and sapphire. High resolution TEM image of the interfaces (d) Al/$Al_2O_3$ (f) Ti/$HfO_2$/AlScN (g) Ti/Au. (h) EDS elemental distribution map near one of the edges of VIA structure.

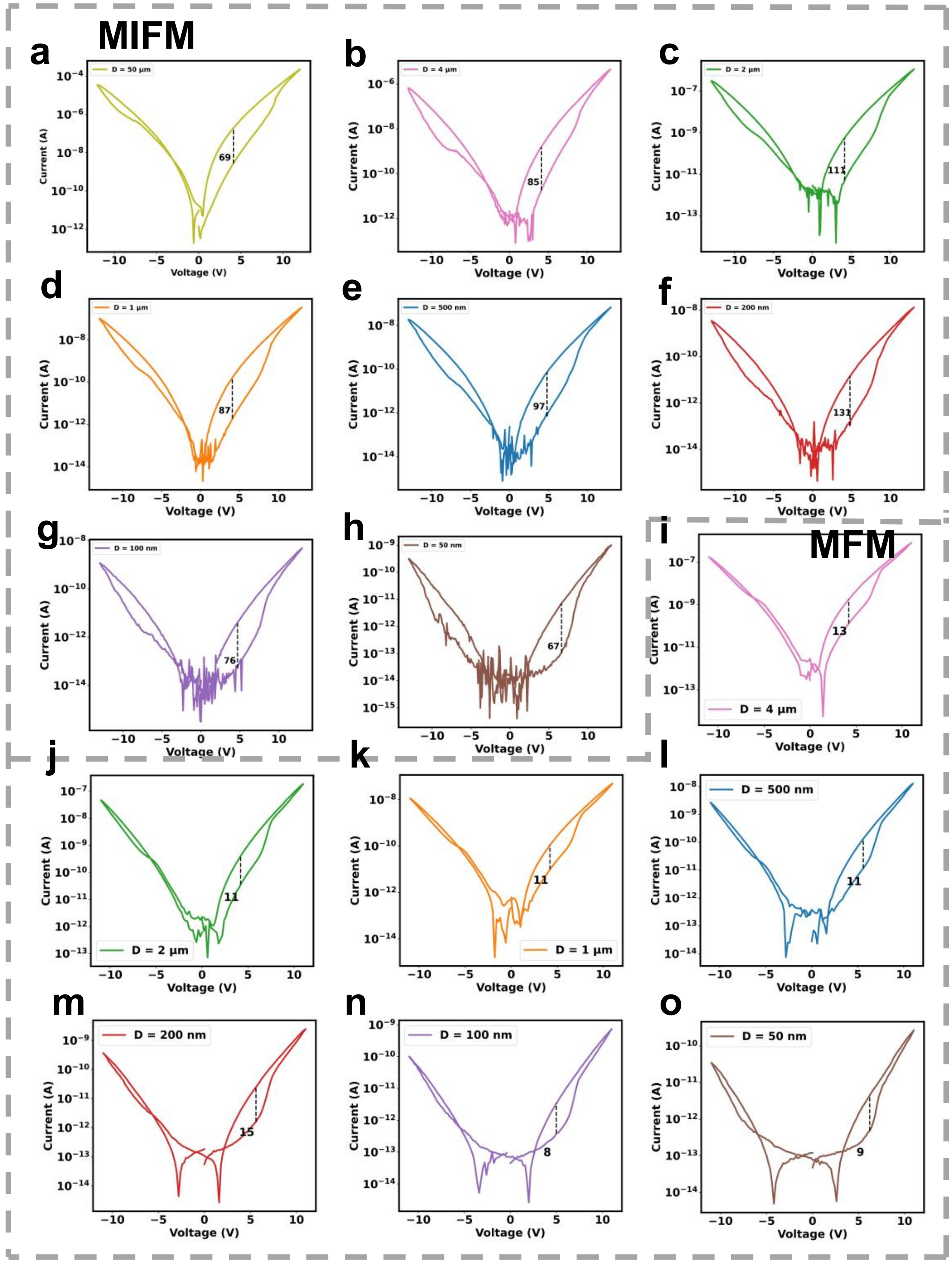

**Figure S4.** Detailed DC current-voltage characteristics for MIFM and MFM structures across different device sizes. (a–h) Individual DC I-V curves for MIFM devices of varying sizes, showing current response under applied voltage, with markers indicating the maximum on-off ratio for positive read voltages. (i–o) Individual DC I-V curves for MFM devices across different sizes,

highlighting their distinct switching characteristics, with maximum on-off ratios also marked for positive read voltages.

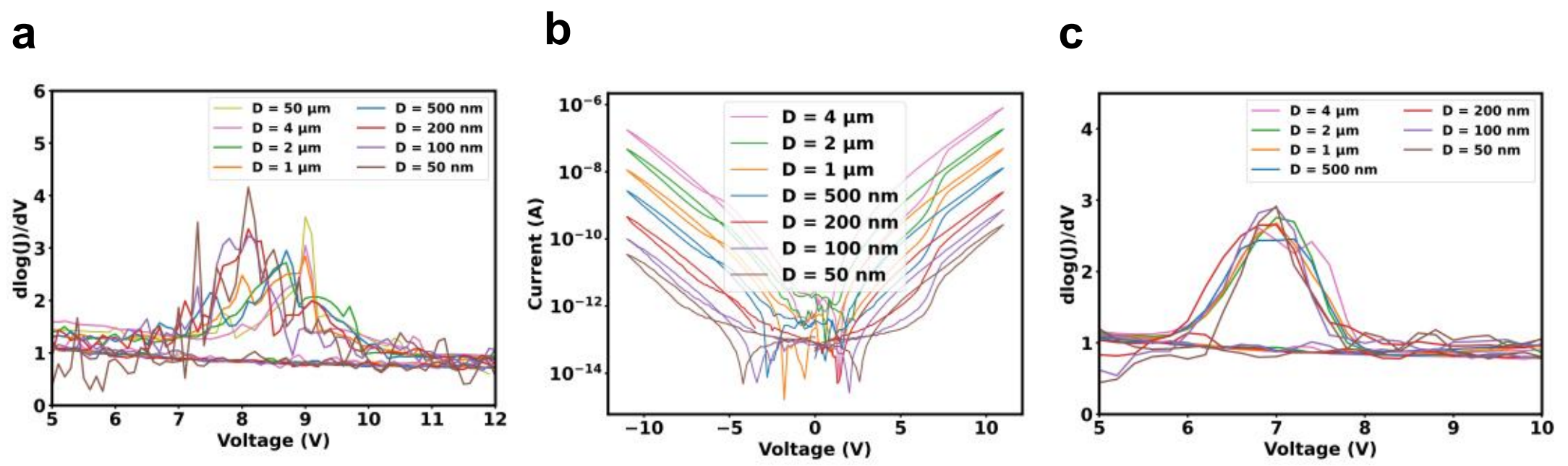

**Figure S5.** DC-IV additional plots. (a) Gradient plot of current density in logarithmic scale as a function of voltage for MIFM devices. (b) Overlay of DC I–V curves for all MFM device sizes, highlighting the impact of scaling on electrical behavior. (c) Gradient plot of current density in logarithmic scale as a function of voltage for MFM devices.

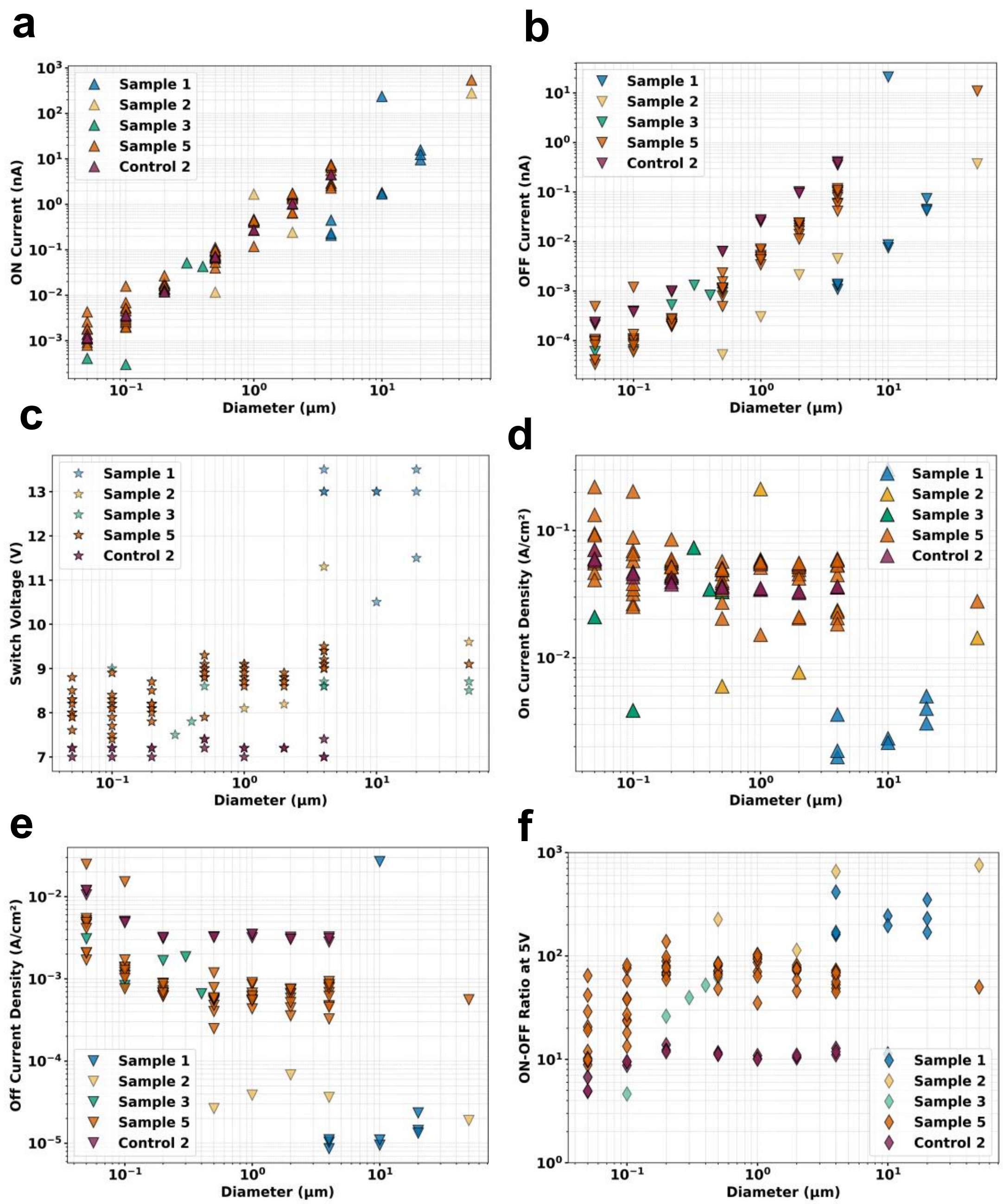

**Figure S6.** Comparison of key electrical parameters across all samples, including (a) on-state current, (b) off-state current, (c) switching voltage, (d, e) current density for each state and (f) on-off ratio at 5V. The dataset includes multiple sample variations: Sample 5 corresponds to the MIFM results presented in the main manuscript, while Control 2 represents the MFM device data. There are slight variations in fabrication processes across the samples: Samples 1 and 2 used PMMA A4 as the mask for etching the $SiO_2$ via, Sample 3 used ZEP520A7 resist, and Sample 5 utilized a ZEP:Anisole 1:1 diluted resist for via etching. Control Sample 2 followed the same process as S5 but omitted the ALD $HfO_2$ deposition. The results highlight the impact of these fabrication variations on device performance.

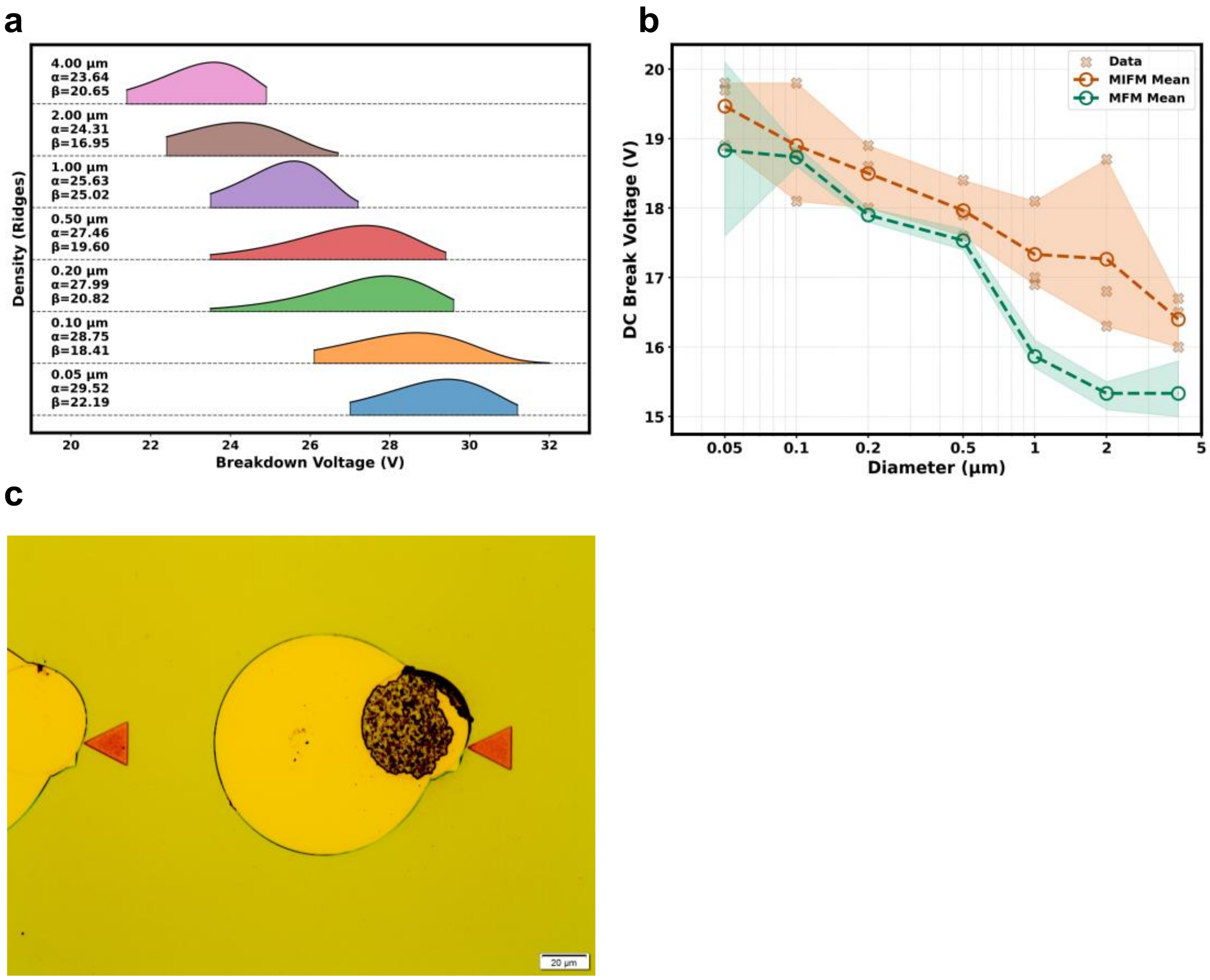

**Figure S7.** Additional insights into breakdown characteristics. (a) Weibull analysis of breakdown voltage for different device sizes, presented as a ridge plot to illustrate statistical distributions. (b) DC breakdown voltage comparison between MIFM and MFM structures, showing the effect of scaling. The MIFM dataset includes 20 devices per size, while the MFM data is averaged over 3 devices per size. (c) Optical Image of a broken device with D = 25 µm.

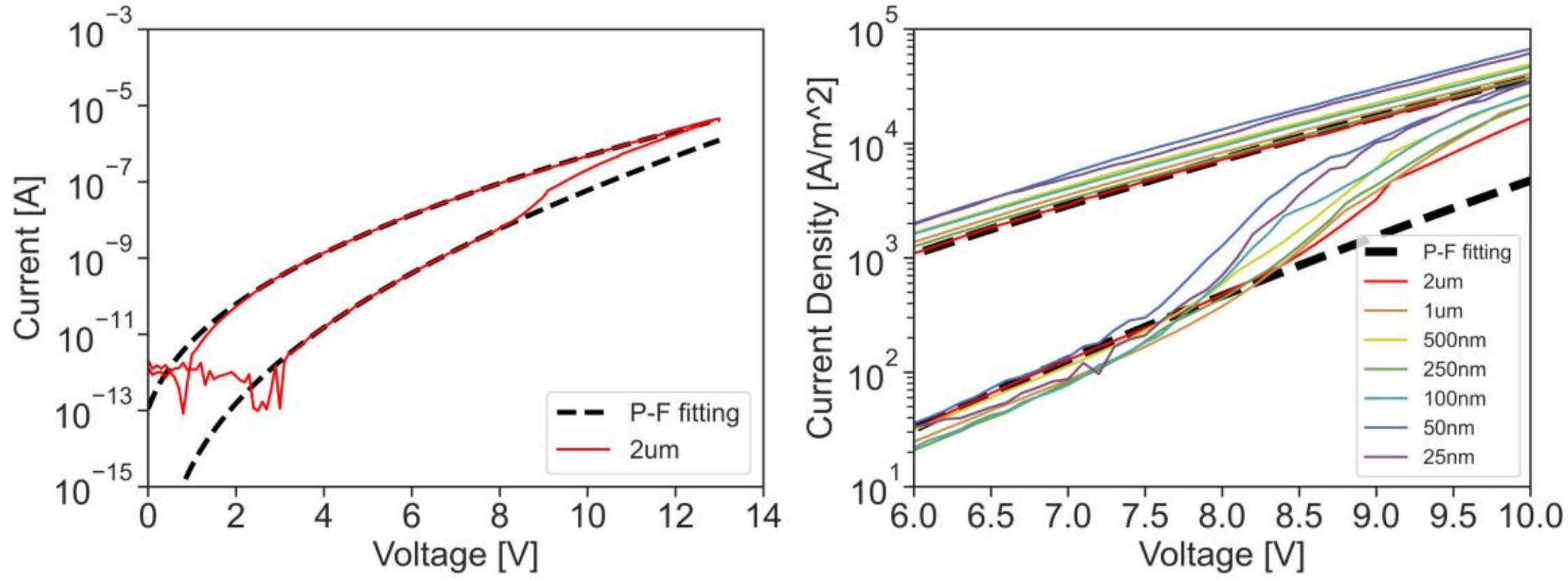

**Figure S8.** Poole-Frenkel (PF) Simulation Model. (a) Fitting of the Poole-Frenkel conduction model in the 5–7 V range for both on-state and off-state, demonstrating agreement with experimental data. (b) Comparison of the current density across different device sizes with the fitted PF model for a 2 μm diameter device. The threshold for ferroelectric switching is observed in the range of 7–7.5 V.

Note 1: PF simulation model

Description for P-F model fitting:

In our simulation, both high resistance state (HRS) and low resistance state (LRS) current is being described by Poole− Frenkel emission due to a relatively high operating voltage of interest (read at 5 V and switching at 7–7.5 V). The current density expression is shown below:

$$J = AE\ exp\left(-\frac{q(\phi - \sqrt{qE/\boldsymbol{\pi}\varepsilon})}{kT}\right)$$

Considering HRS and LRS have opposite polarization $P_r$ and the polarized field $E_p$ within ferroelectric layer is calculated as follow:

$$E_p = \frac{-P_r}{\varepsilon_{r_FE}}\left(1 - \frac{1}{1 + (\frac{\varepsilon_{r_FE}}{\varepsilon_{r_IL}}\frac{t_{FE}}{t_{IL}})}\right)$$

Where $\varepsilon_{r_FE}$ and $\varepsilon_{r_IL}$ are the relative permittivity of ferroelectric layer and insulation layer; $t_{FE}$ and $t_{IL}$ are the thickness of ferroelectric layer and insulation layer, respectively. Thus, we could calculate the effective field $E_{total}$ across ferroelectric layer:

$$E_{ex} = \frac{-V}{t_{FE} + \frac{\varepsilon_{r_{FE}}}{\varepsilon_{r_{IL}}}t_{IL}}$$

$$E_{total} = E_{ex} + E_p$$

Where $V$ is external voltage applied to device from bottom electrode. Bring $E_{total}$ back to Poole–Frenkel emission equation, we could fit the measurement with certain combination of $\phi$ and $A$ for HRS and LRS separately.

In **Figure S8**, DC current sweep data from FeDiode device of 2 μm radius is being fitted with Poole-Frenkel emission in part a. More data from scaling down device (down to 25 nm in radius) is demonstrated in part b, revealing that our material of choice retain a stable conduction mechanism with significantly smaller device area.

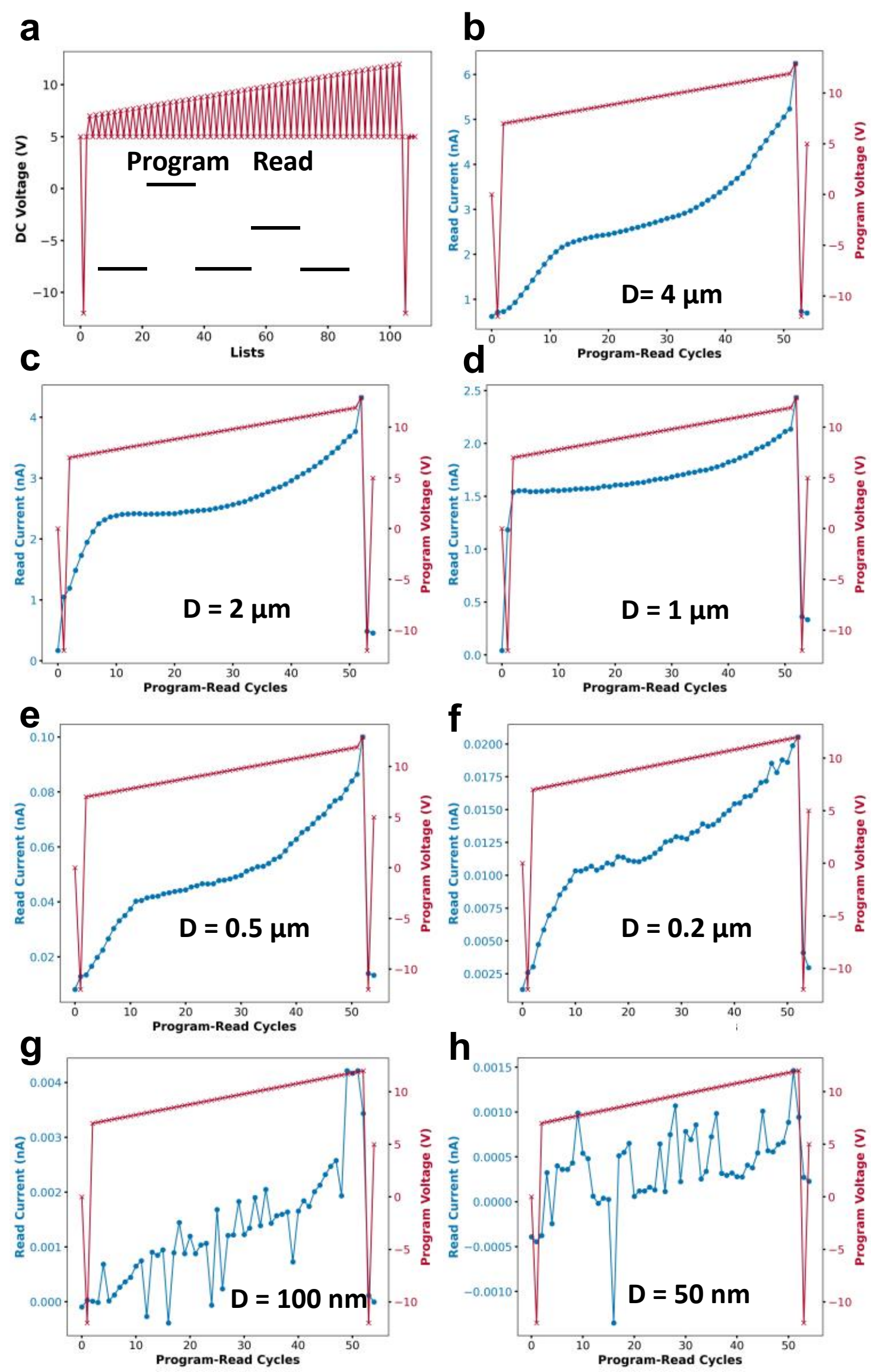

**Figure S9.** Multistate Behavior Under Fine-Step Programming. (a) DC voltage sequence applied to the device, starting from a fully erased state and progressively programmed with 0.1 V increments per cycle. (b–h) Read current response as a function of program-read cycles for different device sizes, ranging from 4 μm (b) down to 50 nm (h), illustrating size-dependent multistate characteristics and stability. For device radius smaller than 100 nm, the accuracy of signal decreases.

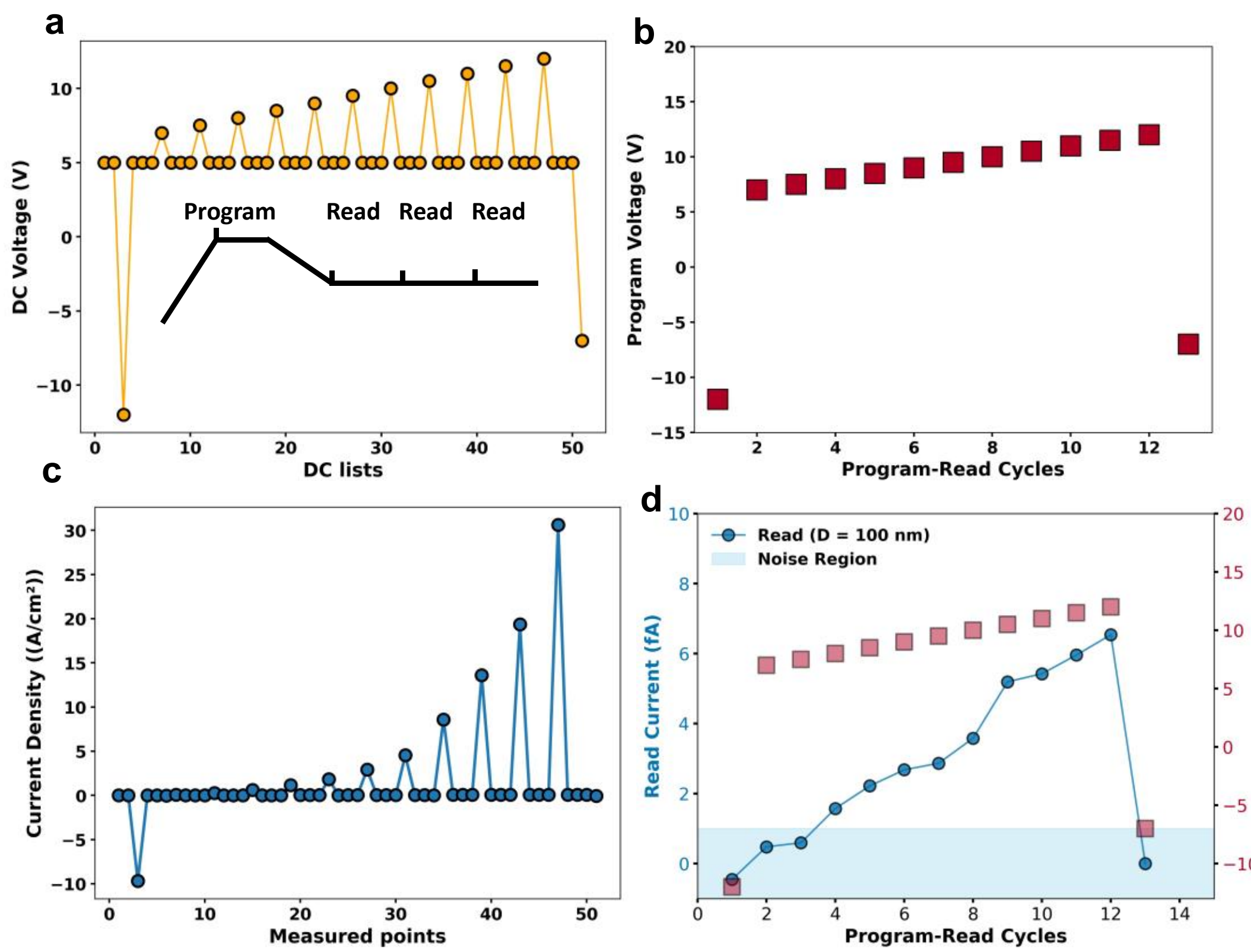

**Figure S10.** Illustration of the program-read multistate testing procedure. (a) Full DC test sequence consisting of multiple program-read-read-read cycles with incrementally increasing program voltage. The read operation is repeated three times to ensure signal accuracy. (b) Extracted program voltage values from the test sequence. (c) Current response over the complete DC test sequence, showing the effect of increasing program voltage. (d) Filtered read current response at each program voltage, with the shaded region indicating the measurement noise limit.

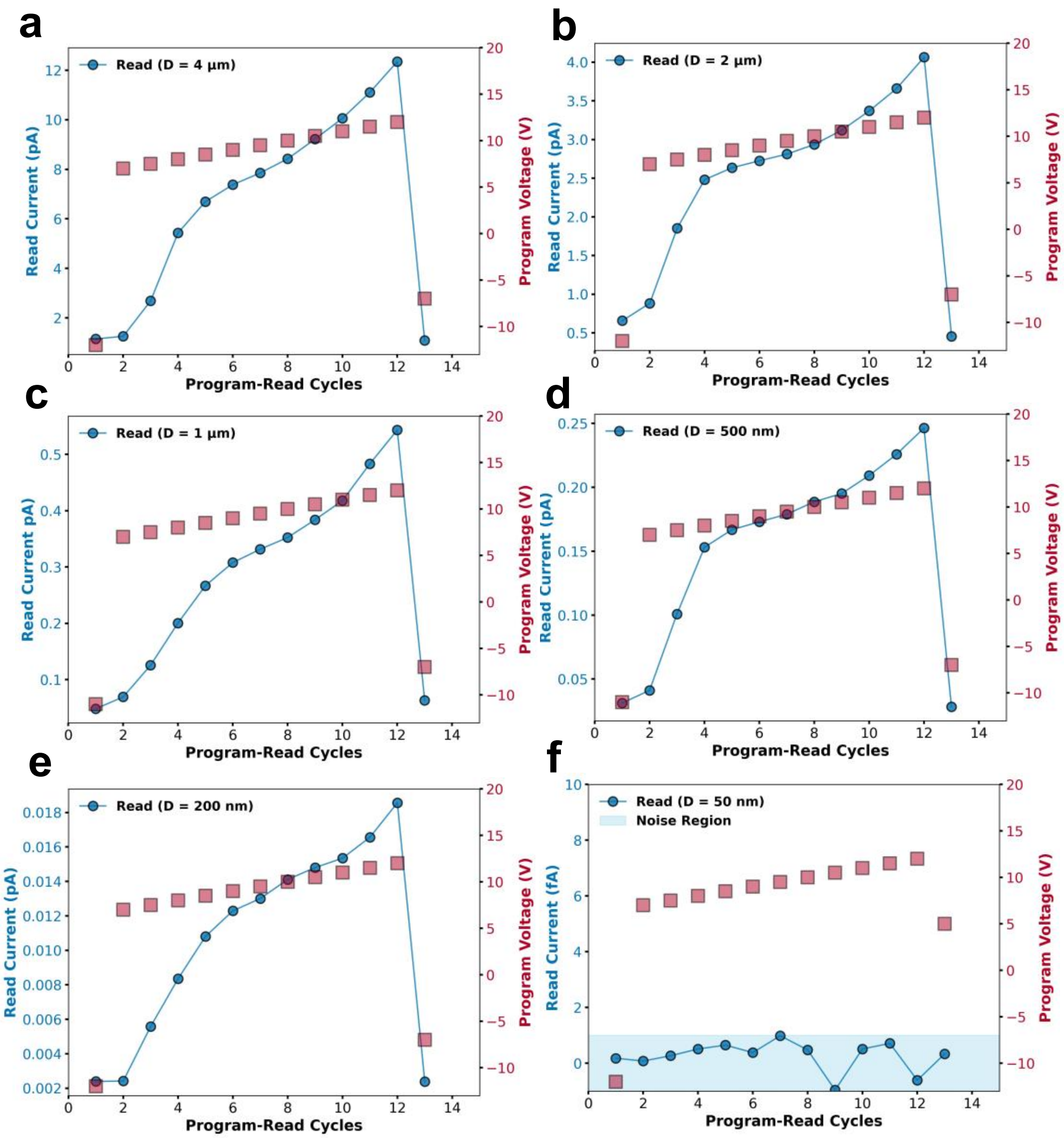

**Figure S11**. Current response of devices with different diameters under program-read cycles. (a–e) Read current evolution for devices with diameters of 4 μm, 2 μm, 1 μm, 500 nm, and 200 nm, respectively, showing progressive changes in read current as program cycles increase. (f) Read current response for the 50 nm diameter device, where signal detection is constrained by measurement noise, as indicated by the shaded region.

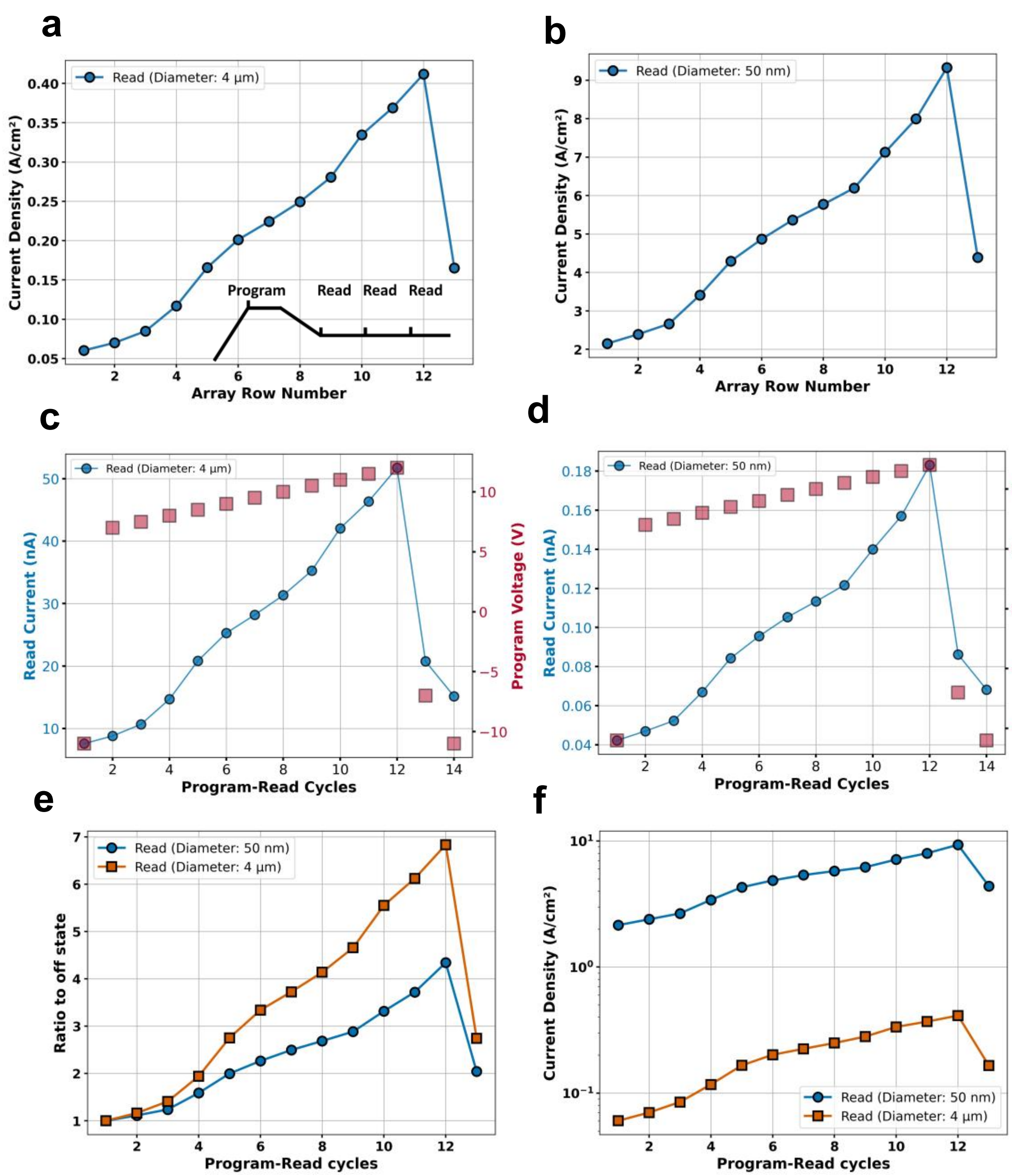

**Figure S12.** Multistate test results read at 7V under program-read cycles. Current density evolution for a (a) 4 μm device (b) 50 nm device. Read current and corresponding program voltage across program-read cycles for the (c) 4 μm device (d) 50 nm device. (e) Comparison of the ratio to the off-state for 4 μm and 50 nm devices, showing differences in scaling behavior.(f) Comparison of current density between 4 μm and 50 nm devices, highlighting size-dependent variations in multistate retention.

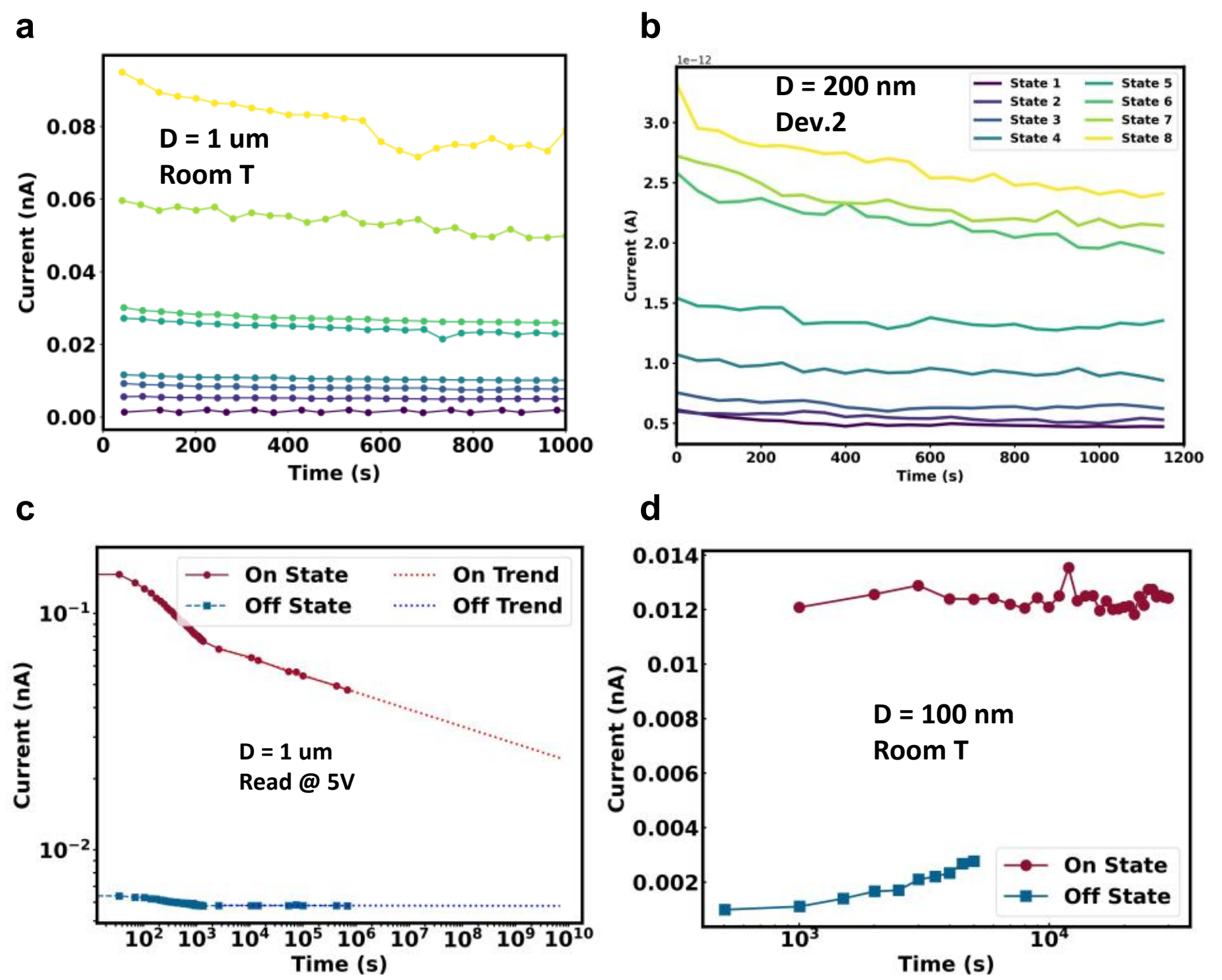

**Figure S13**. Retention Test Across Different Device Sizes. (a) Multistate retention behavior over 1000 seconds for a device with a 1 μm diameter read at 5V. (b) Multistate retention behavior over 1000 seconds for a device with a 200 nm diameter read at 5V. (c) Retention measurement extended to 5*10^5 seconds for a 1 μm diameter device read at 5V. (d) Retention measurement extended to 5*10^4 seconds for a 100 nm diameter device read at 5V.

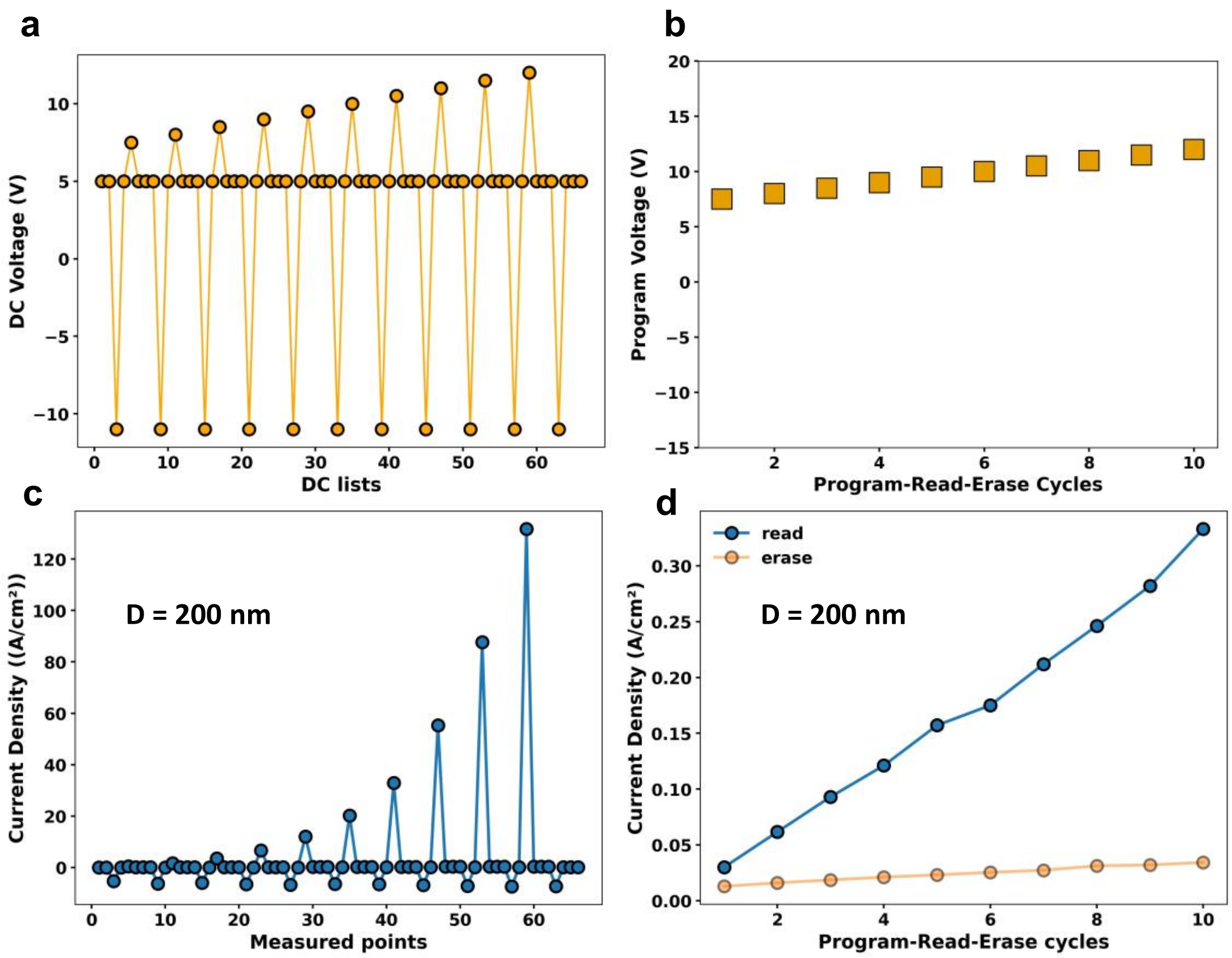

**Figure S14**. Multistate test under program-read-erase cycles. (a) DC voltage sequence applied during the test, illustrating the alternating program and erase operations. (b) Evolution of program voltage over multiple program-read-erase cycles. (c) Complete current response with Diameter of 200 nm across the full sequence, showing dynamic changes in current density. (d) Comparison of read current and erase current across cycles, highlighting the stability of the erased state and progressive increase in read current.

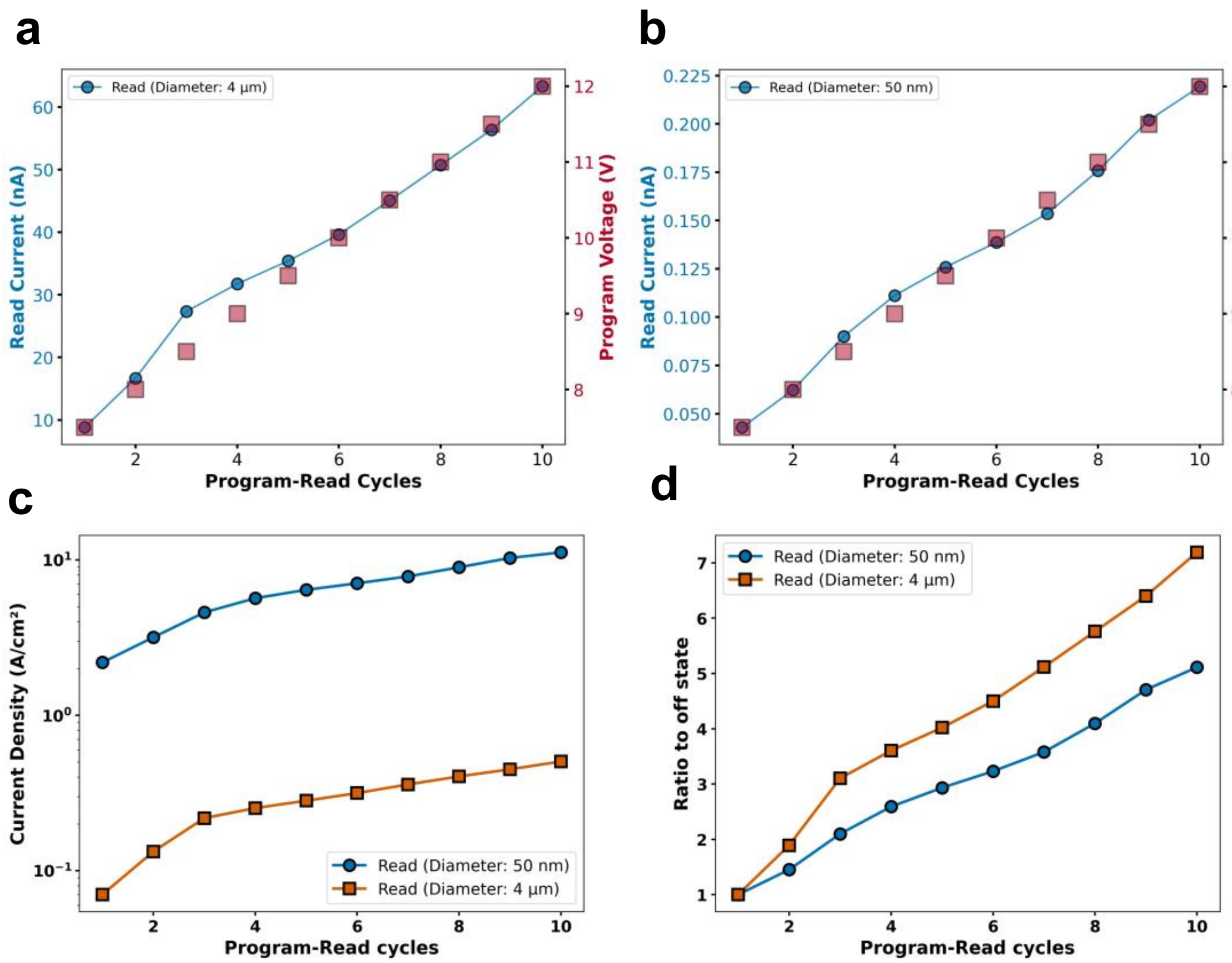

**Figure S15.** Multistate test results read at 7V under program-read-erase-read cycles. (a) Read current and corresponding program voltage across program-read cycles for the 4 μm device. (b) Read current and corresponding program voltage across program-read cycles for the 50 nm device. (c) Comparison of current density between 4 μm and 50 nm device. (d) Comparison of the ratio to the off-state for 4 μm and 50 nm devices, showing similarity in scaling behavior.

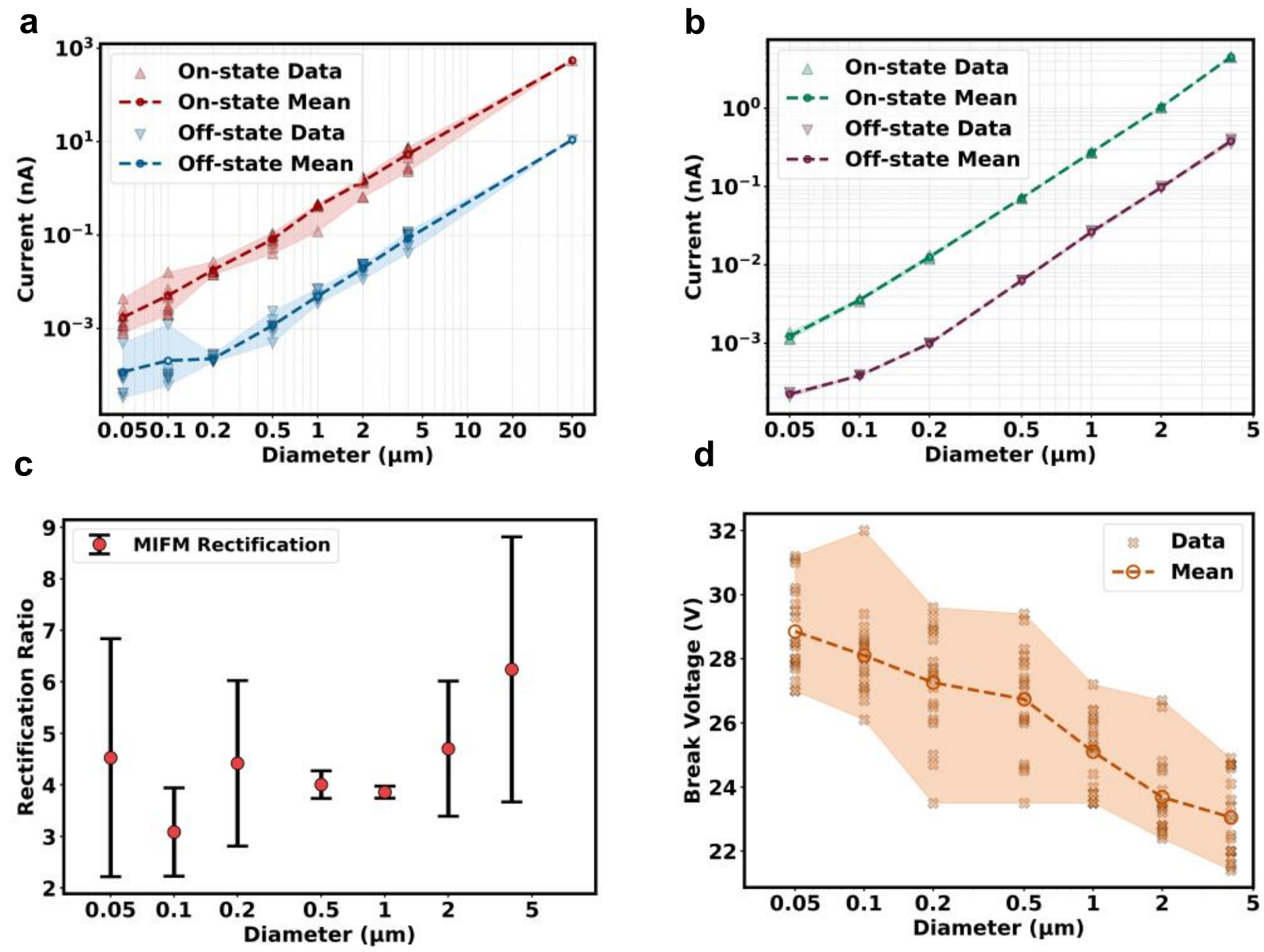

**Figure S16.** Statistical analysis of device performance. (a) On-state and off-state currents measured at 5 V for MIFM structures, showing well-separated resistive states across device diameters. (b) On-state and off-state currents measured at 5 V for MFM structures. (c) Rectification ratio of MIFM devices, calculated as the absolute current ratio at +5 V and –5 V. (d) Breakdown voltage distribution of MIFM devices measured using a 10 kHz triangular waveform, showing individual device values and averaged trend.

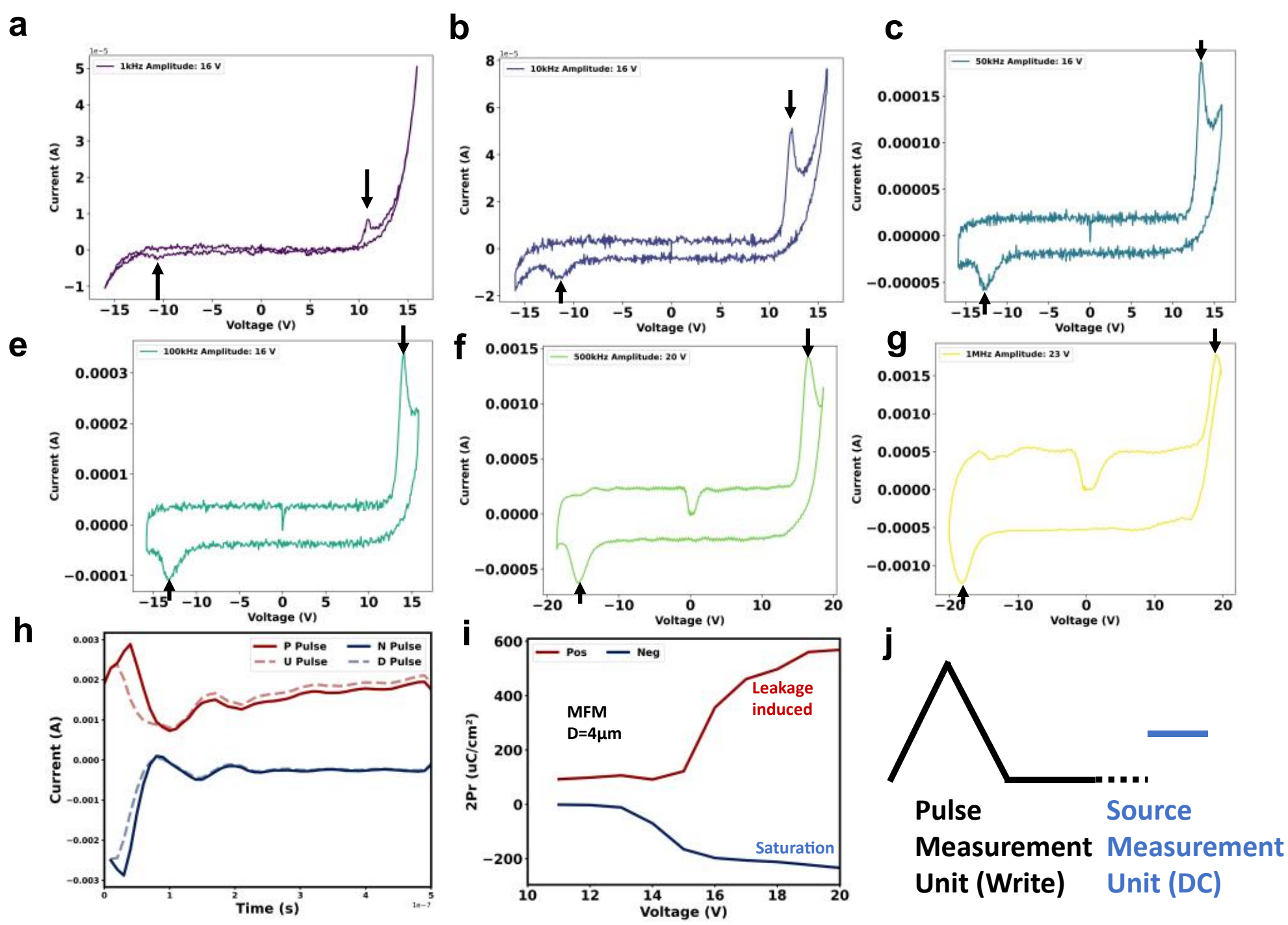

**Figure S17.** Ferroelectric switching behavior under triangular wave and pulse measurements. (a–g) Current–voltage (I–V) characteristics of MFM structures under triangular voltage pulses of varying frequency and amplitude. The ferroelectric switching peaks are marked with black arrows. Increased baseline current with frequency is attributed to capacitive charging of via pads. (h) Polarization current transients from PUND measurements showing clear separation between switched (P/N) and unswitched (U/D) states. (i) Extracted 2Pr from pulsed measurements on a 4 µm MFM device, showing saturation of negative polarization and leakage-induced increase in the positive branch at high voltage. (j) Testing schematics for triangular AC-write-DC-read.

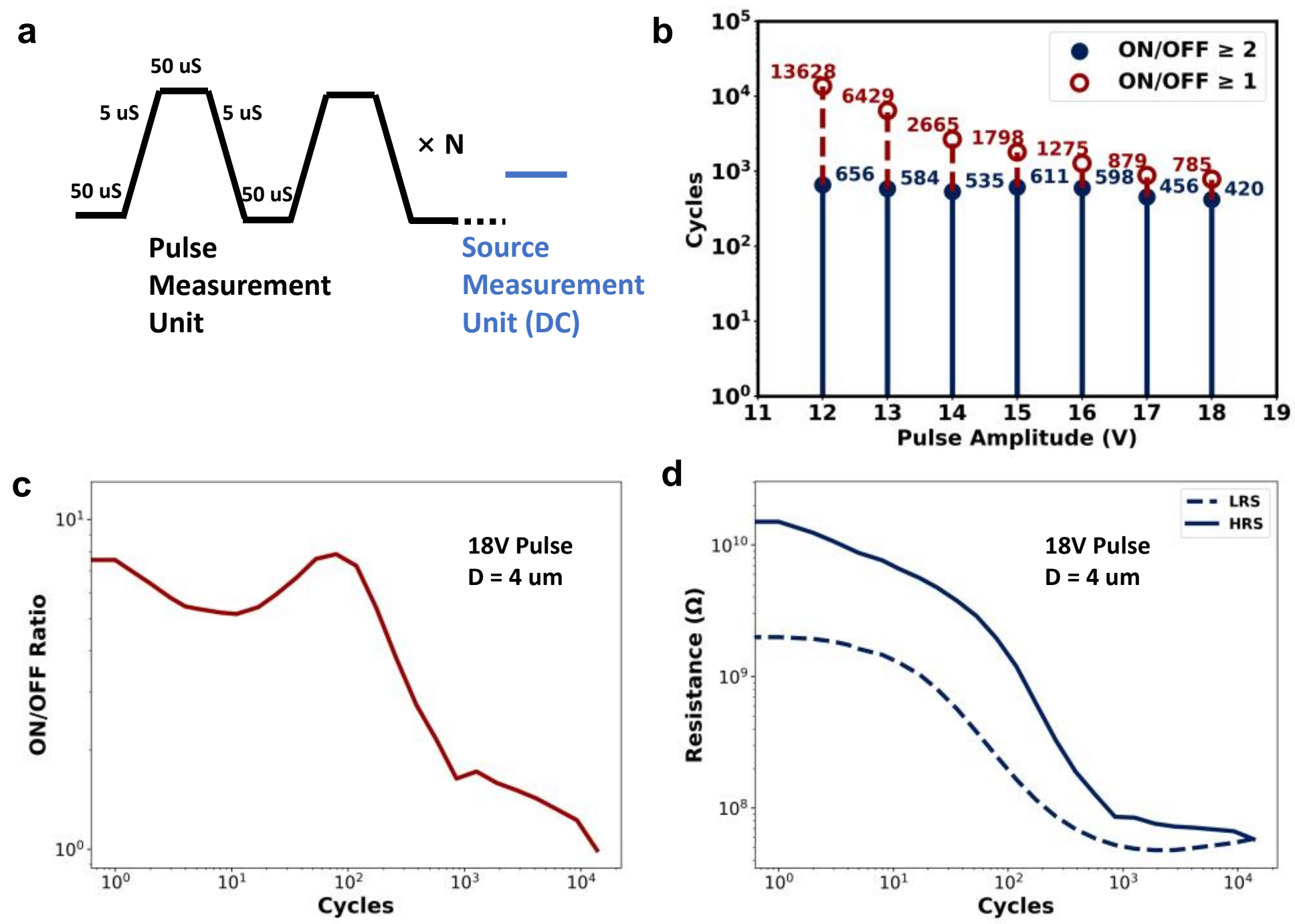

**Figure S18.** Pulse-programmed endurance characterization of MIFM devices. (a) Schematic of the test protocol: a sequence of ±18 V pulses with 50 µs width and 5 µs rise/fall time is applied using the Pulse Measurement Unit (PMU), followed by low-noise DC readout using the Source Measurement Unit (SMU). (b) Summary of endurance results for MIFM devices (D = 4 µm) under varying pulse amplitudes, with endurance cycle counts labeled. Endurance is defined by the number of cycles before ON/OFF ratio drops below 2 (blue) or 1 (red). (c) Evolution of ON/OFF ratio over pulse cycles at 18 V amplitude. (d) Corresponding resistance states (LRS and HRS) extracted from DC readout at each endurance cycle, showing a clear degradation trend.

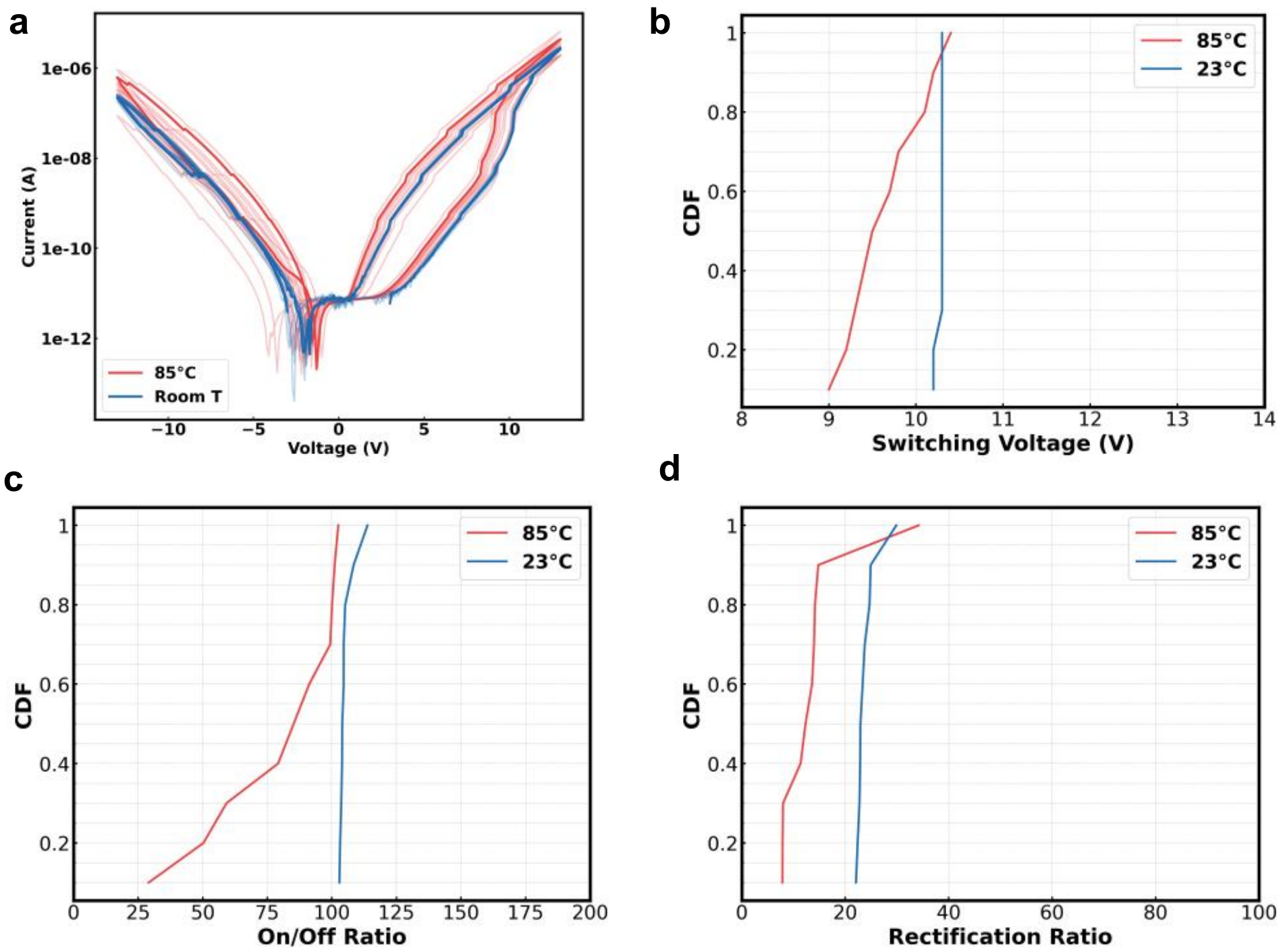

**Figure S19.** Impact of elevated temperature on DC characteristics of MIFM devices (D = 2 µm). (a) DC current–voltage curves measured at room temperature (23 °C, blue) and 85 °C (red) for ten representative devices. The consistent switching behavior and stable ON/OFF characteristics across both temperatures confirm that the resistive switching originates from ferroelectric polarization, rather than thermally activated effects such as Joule heating or oxide breakdown. (b) Cumulative distribution function (CDF) of switching voltages extracted from I–V hysteresis measurements. (c) CDF of ON/OFF ratio measured at ±5 V. (d) CDF of rectification ratio, defined as the absolute current ratio between forward and reverse biases at 5 V. The results highlight temperature-dependent variations in leakage current and nonlinearity while preserving polarization-driven switching contrast.

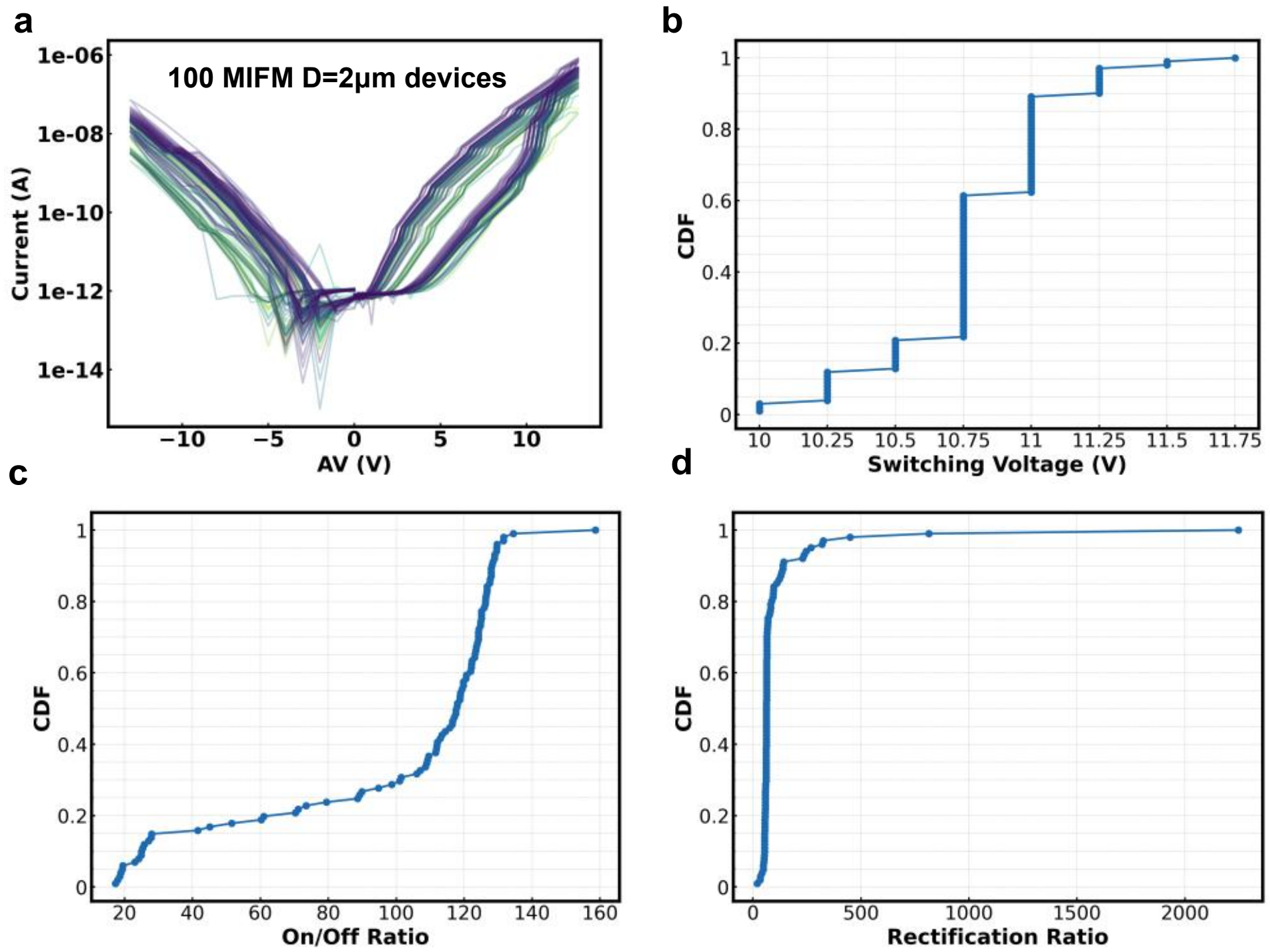

**Figure S20.** Device-to-device variability characterization across 100 MIFM devices with 2 μm diameter. (a) DC IV curves of all devices, showing consistent resistive switching behavior. (b) CDF of switching voltage. The slightly elevated switching voltages are attributed to chip-to-chip variation. (c) CDF of ON/OFF ratio at ±5 V, illustrating statistical dispersion in resistive states. (d) CDF of rectification ratio, defined as the absolute forward/reverse current ratio at 5 V. These results highlight robust performance reproducibility and statistical uniformity across a large device population.

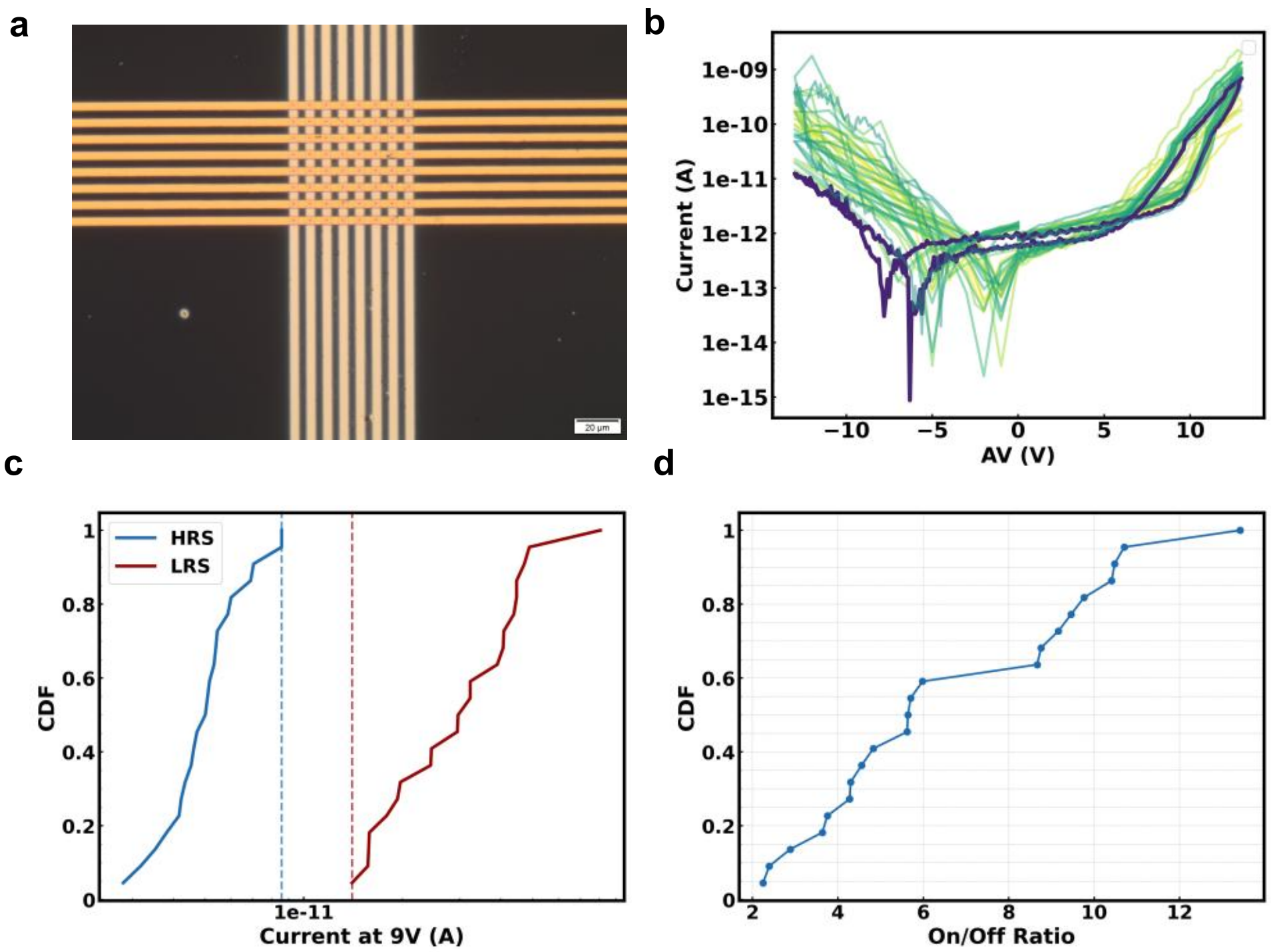

**Figure S21.** 8×8 crossbar demonstration of selector-free MIFM device integration.
(a) Optical micrograph of an 8×8 crossbar array comprising 1 µm diameter MIFM devices. Each metal line is 4 µm wide with a 4 µm pitch. (b) DC current–voltage (I–V) characteristics from 22 devices measured within a single crossbar, showing consistent resistive switching behavior. (c) CDF of high-resistance state (HRS) and low-resistance state (LRS) current values at 9 V, illustrating a clear sense margin between the two states. (d) CDF of ON/OFF ratio, demonstrating distinguishable resistance states. Further performance optimization is possible with improved fabrication uniformity.

Note 2: Storage density calculation:

$$\text{Storage density} = \frac{1}{\mathbf{4F^2}}$$

For binary operation:

$$\text{Storage density} = 1\,b \times \frac{1}{\mathbf{4} \times (40 \times 10^{-9} m)^2} \times 10^{-9}\, {}^{Gb}/_{b} \times 10^{-6}\, {}^{m^2}/_{mm^2}$$
$$= 0.225\ {}^{Gb}/_{mm^2}$$

For 3 bits operation:

$$\text{Storage density} = 3\,b \times \frac{1}{\mathbf{4} \times (40 \times 10^{-9} m)^2} \times 10^{-9}\, {}^{Gb}/_{b} \times 10^{-6}\, {}^{m^2}/_{mm^2}$$
$$= 0.675\ {}^{Gb}/_{mm^2}$$

Note 3: Crossbar integration feasibility.

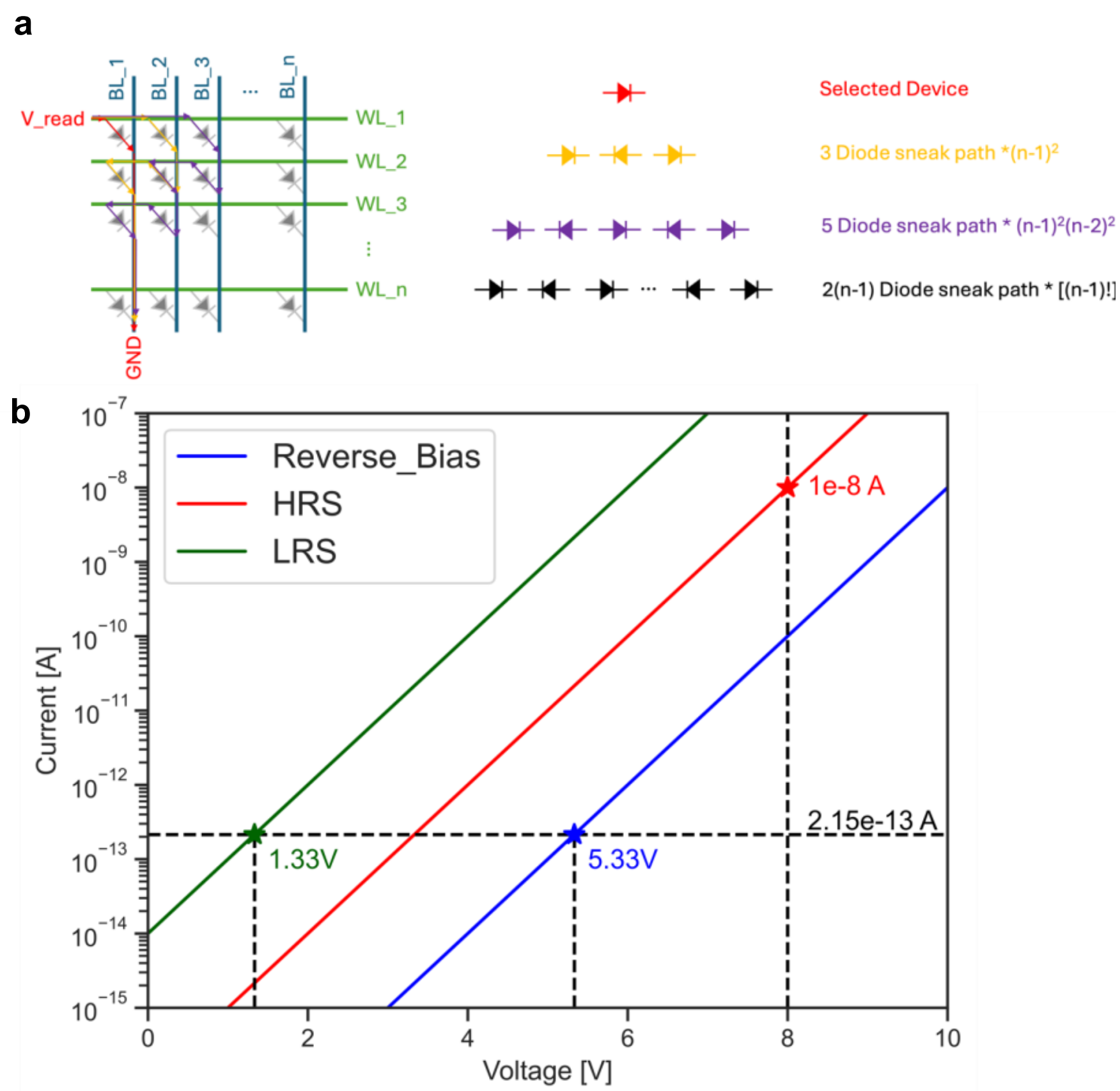

**Figure S22.** Selector-free readout feasibility enabled by diode-like rectification. (a) Schematic of sneak path configurations in an n × n crossbar array. The selected device (red) is read through the intended path, while potential leakage through unselected devices occurs via various sneak paths, including 3-diode $[(n-1)^2]$, 5-diode $[(n-1)^2(n-2)]$, and 2(n–1)-diode paths, depending on the array topology. (b) Simulated current–voltage characteristics of LRS, HRS, and reverse-biased diode states, extracted from experimental measurements of 4 µm MIFM devices. At a read current threshold of $10^{-8}$ A, reverse-biased devices limit leakage below ~$2.15\times10^{-13}$ A. The high rectification and current contrast between read and sneak paths demonstrate the feasibility of selector-free operation.

For an $n \times n$ crossbar array, there are $(n-1)$ distinct variants of sneak paths, each composed of $2(n-1)$ FeDiodes—$(n-2)$ of which are under reverse bias, with the remainder under forward bias. The total number of sneak paths scales quadratically with the array size, while their effective contribution diminishes due to the nonlinear and rectifying nature of the FeDiode.

As an example, in an 8 × 8 array, 3-diode sneak paths are analyzed under a worst-case assumption where the selected device is in the high-resistance state (HRS), and all other devices are in the low-resistance state (LRS). Assuming the device DC I–V characteristics follow a logarithmic trend with 100× separation between LRS, HRS, and reverse-bias states, and that the 8 V read voltage drops across a forward–reverse–forward configuration, each sneak path contributes ~$2.15 \times 10^{-13}$ A. With 49 such paths active, the total sneak current remains several orders of magnitude lower than the HRS read current (~$1 \times 10^{-8}$ A at 8 V). This analysis confirms that the intrinsic nonlinearity and rectification of FeDiodes effectively suppress sneak-path leakage, enabling reliable selector-free readout.